\documentclass{aastex63}

\begin{document}

\title{Cataclysmic Variables in the Second Year of the Zwicky Transient Facility}

\correspondingauthor{Paula Szkody}
\email{szkody@astro.washington.edu}

\author[0000-0003-4373-7777]{Paula Szkody}
\affil{University of Washington,
Department of Astronomy, Box 351580,
Seattle, WA 98195, USA}

\author{Claire Olde Loohuis}
\affil{University of Washington,
Department of Astronomy, Box 351580,
Seattle, WA 98195, USA}

\author[0000-0001-5530-2872]{Brad Koplitz}
\affil{University of Washington,
Department of Astronomy, Box 351580,
Seattle, WA 98195, USA}

\author[0000-0002-2626-2872]{Jan van Roestel}
\affiliation{Division of Physics, Mathematics and Astronomy,
California Institute of Technology,
Pasadena, CA 91125, USA}

\author[0000-0002-8257-9727]{Brooke Dicenzo}
\affiliation{University of Washington,
Department of Astronomy, Box 351580,
Seattle, WA 98195, USA}

\author[0000-0002-9017-3567]{Anna Y. Q.~Ho}
\affiliation{Division of Physics, Mathematics and Astronomy,
California Institute of Technology,
Pasadena, CA 91125, USA}
\affiliation{Department of Astronomy, University of California, Berkeley, 501 Campbell Hall, Berkeley, CA, 94720, USA}
\affiliation{Miller Institute for Basic Research in Science, 468 Donner Lab, Berkeley, CA 94720, USA}

\author{Lynne A. Hillenbrand}
\affiliation{Division of Physics, Mathematics and Astronomy,
California Institute of Technology,
Pasadena, CA 91125, USA}

\author[0000-0001-8018-5348]{Eric C. Bellm}
\affiliation{University of Washington,
Department of Astronomy, Box 351580,
Seattle, WA 98195, USA}






\author{Richard Dekany}
\affiliation{Caltech Optical Observatories, California Institute of Technology, Pasadena, CA 91125, USA}

\author{Andrew J. Drake}
\affiliation{Division of Physics, Mathematics and Astronomy, California Institute of Technology,
Pasadena, CA 91125, USA}

\author[0000-0001-5060-8733]{Dmitry A. Duev}
\affiliation{Division of Physics, Mathematics, and Astronomy, California Institute of Technology, Pasadena, CA 91125, USA}

\author{Matthew J. Graham}
\affiliation{Division of Physics, Mathematics and Astronomy, California Institute of Technology,
Pasadena, CA 91125, USA}



\author{Mansi M. Kasliwal}
\affiliation{Division of Physics, Mathematics and Astronomy, California Institute of Technology,
Pasadena, CA 91125, USA}


\author[0000-0003-2242-0244]{Ashish~A.~Mahabal}
\affiliation{Division of Physics, Mathematics and Astronomy, California 
Institute of Technology, Pasadena, CA 91125, USA}
\affiliation{Center for Data Driven Discovery, California Institute of 
Technology, Pasadena, CA 91125, USA}

\author[0000-0002-8532-9395]{Frank J. Masci}
\affiliation{IPAC, California Institute of Technology, 1200 E. California Blvd, Pasadena, CA 91125, USA}

\author{James D. Neill}
\affiliation{Division of Physics, Mathematics and
Astronomy, California Institute of Technology,
Pasadena, CA 91125, USA}



\author{Reed Riddle}
\affiliation{Division of Physics, Mathematics and Astronomy, California Institute of Technology,
Pasadena, CA 91125, USA}

\author[0000-0001-7648-4142]{Benjamin Rusholme}
\affiliation{IPAC, California Institute of Technology, 1200 E. California Blvd, Pasadena, CA 91125, USA}

\author{Jesper Sollerman}                        
\affiliation{The Oskar Klein Centre, Department of Astronomy, 
AlbaNova, SE-106 91 Stockholm, Sweden}

\author{Richard Walters}
\affiliation{Caltech Optical Observatories, California Institute of Technology, Pasadena, CA 91125, USA}
 

\begin{abstract}
Using a filter in the GROWTH Marshal based on color and the amplitude and the timescale of variability, we have identified
372 objects as known or candidate cataclysmic variables (CVs) during the second year of operation of the Zwicky Transient Facility (ZTF). From the available difference imaging data, we found that 93 are previously confirmed CVs, and 279 are strong candidates. Spectra of four of the candidates confirm them as CVs by the presence
of Balmer emission lines, while one of the four has prominent HeII lines indicative of containing a magnetic
white dwarf. Gaia EDR3 parallaxes are available for 154 of these systems, resulting in distances from 108-2096 pc and
absolute magnitudes in the range of 7.5-15.0, with the largest number of candidates between 10.5-12.5. The total numbers are 21\% higher than from the previous year of the survey with a greater number
of distances available but a smaller percentage of systems close to the Galactic plane. Comparison of these findings with a machine learning method of searching all the light curves
reveals large differences in each dataset related to the parameters involved in the search process.

\end{abstract}


\section{Introduction} 
The Zwicky Transient Facility (ZTF) is a three year photometric survey that uses a wide 47 deg$^{2}$ field of view camera on the Palomar 48-inch telescope with $g,r,i$ filters
 \citep{B19a,B19b,G19,M19,D20,Z16}. During the first two years of the survey, the 40\% of the public time was used to observe the available sky every three nights in $g$ and $r$ filters, and the Galactic plane every night, while the rest of the time was divided between programs designed by the partnership (40\%) and Caltech (20\%). Besides its wide field and depth of coverage, one of the primary advantages of ZTF over other surveys such as ASASSN and Gaia is the increased temporal observations of the Galactic plane, especially within the partnership portion. The official survey began on 2018 March 18 and has been producing nightly alerts on transient/variable phenomena, as well as public and partnership data releases that can be accessed through IPAC\footnote{https://irsa.ipac.caltech.edu/Missions/ztf.html}. The first data release (DR1) took place on 2019 May 8, the second (DR2) on 2019 December 11, the third (DR3) on 2020 June 24, the fourth (DR4) on 2020 December 9 and the fifth (DR5) on 2021 March 31.

This paper is the second in a series identifying cataclysmic variables (CVs) from ZTF public and partnership data using the GROWTH Marshal\citep{K19} to filter the alerts during the interval from 2019 June 1 and 2020 May 31. The first paper\citep{Sz20} presented the software filter used in the Marshal (based on point source, $g-r$ color $\le$0.6 and magnitude change $\ge$2 mag within a timescale of $\le$2 days) and provided a list of 90 previously confirmed CVs and 218 strong candidates found in the ZTF alerts from June 2018 until May 31 2019. These objects were found based on the shape and colors of their light curves, and 29 of the candidates were confirmed by obtaining spectra. 

 \citet{W95} provides an overall review of the different types of CVs that are being found
 in the ZTF data. They are all close binaries with mass transfer from a companion (usually a late main sequence star) to a white dwarf.
 The main type being discovered in sky surveys is dwarf novae as they are easily located by the brightness changes during a disk instability outburst. A few novalikes are found when they undergo low and high accretion state changes. To further confirm the candidates and refine the classifications, spectra are needed. The presence of prominent hydrogen Balmer emission lines confirm a dwarf nova or novalike CV, while helium lines confirm an AM CVn or a novalike system containing a magnetic white dwarf (polar or intermediate polar), or a system with a very high accretion rate (SW Sex star). As the list of confirmed CVs grows and contains astrometry (from Gaia), the results can be used to test population models of close binary evolution \citep{HRP97,Mc19}.

\section{Identifying CVs} 

Each night from 2019 June 1 to 2020 May 31 (except for the month of 2020 March when the system was down due to repairs of the filter exchanger), the light curves of the candidates created from the alerts that passed the Marshal CV filter that night (based on the point source, color, and magnitude change listed above) were examined. These light curves provided a 30 day interval of observation prior to the night requested. The candidates were then saved
if the light curve appeared to result from a dwarf nova outburst or a change in the accretion state of a novalike system. The saved systems then accumulated further data if obtained, allowing for a later classification.
The saved candidates were then cross-checked with other catalogs, including SIMBAD \citep{W20}, the AAVSO VSX catalog \citep{W07}, the Sloan
Digital Sky Survey \citep{Y00}, the Catalina Real-time
Transient Survey (CRTS) \citep{D09,D14}, MASTER \citep{L10} and
ASAS-SN \citep{Sh14} to see if they were previously known or candidate CVs. 

While spectra are the ultimate confirmation that a candidate is a CV, various circumstances in the past year (the loss of blue capability in the Apache Point Observatory spectrograph and telescope shutdowns from the pandemic) prevented obtaining the same numbers of spectra as in Paper I. Only four confirmation spectra showing Balmer emission lines are
available from ZTF accessible facilities, two using the
Low Resolution Imaging Spectrometer (LRIS) \citep{O95} on the Keck telescope, one with the Floyds spectrograph on the Las Cumbres 2m telescope at Haleakala \citep{Br13} and one from the SPRAT at the Liverpool telescope \citep{St04}. These spectra are discussed in detail below. 

\section{Results}

The scans of the usable nights from the GROWTH Marshal with the CV filter
yielded 93 previously confirmed CVs (generally from spectra but in a few cases from the presence of a superhump outburst feature in the light curve
\citep{W95}), and 279 strong candidates based on their ZTF light curves.  Table 1 provides a list of
the previously confirmed objects, and Table 2 lists the
strong candidates.  Some sources are also listed as candidates in CRTS or MASTER, but if they have not been confirmed with spectra, we placed them in Table 2. As in Paper I, we
will refer to the objects by the use of an abbreviated
RA(HHMM) and Dec(Deg) i.e. ZTF0014+59, with the full coordinates given in the tables.  Also included in the tables are the Galactic latitude, the observed ZTF range in magnitudes from outburst peak to quiescence, or from high to low accretion states, the DR3 Gaia parallax and errors in mas (for measurements more than 3 times the error), the distances in parsecs  (simply using the inverted parallax), the absolute
magnitude at the ZTF observed minimum magnitude, the
number of normal outbursts and longer superoutbursts (SOBs) observed in the Marshal light curves, the number of days of ZTF coverage available between 2019 June 1 and 2020 May 31, if photometry of the source exists in
the Sloan Digital Sky Survey (SDSS) footprint or in the CRTS, if any spectra were obtained with the ZTF instruments (Table 2) or available from the SDSS or the 
literature (Table 1), and any other relevant information.

Figure 1 shows a few examples of the different types of light curves (an SOB, normal short-cycle outbursts, high/low states) 
that led to the classification as a CV candidate in Table 2.

\subsection{Spectroscopic Confirmations}
Only four objects from Table 2 were able to be confirmed from the presence of Balmer emission lines (Figures 2 and 3). While the Spectral Energy Distribution Machine \citep{Bl18} on the Palomar 60-inch telescope obtained several spectra, they were observed near outburst and
only showed a blue continuum from the disk with no emission visible. The medium resolution Keck spectra of the two objects with light curves in the bottom row of Figure 1 (ZTF2134-02 and ZTF2131+49) are shown in Figure 2. These objects were obtained at quiescence and both reveal strong Balmer emission, while ZTF2131+49 also has strong helium lines, especially \ion{He}{2}4686. Thus, this object is a candidate for a system containing a magnetic white dwarf and is worth further followup. The lower resolution spectra from the LCO and SPRAT spectrographs for ZTF0618+22 and ZTF1928+55 were observed near outburst but do show the presence of Balmer emission, confirming them as CVs.

\subsection{The Galactic plane}

As shown in Paper I, the ZTF inclusion of the galactic plane in its footprint results in more new candidates in this area of the sky, compared to the known
candidates. This is further confirmed with the second year data, although with
smaller differences, likely due to changes in the portion of the public survey time spent on
the plane (some of the nightly plane coverage was shifted to coverage of TESS fields). The left
panel of Figure 4 shows the number of known systems (Table 1) and candidates
(Table 2) while the right panel compares the first and second years of data
on all objects. While 18\% of the known systems are within 10$^{\circ}$ of
the plane, 25\% of the candidates are within this range. This compares to 23\% and 45\% in the first year.

\subsection{Absolute Magnitudes}

The EDR3 Gaia parallaxes \citep{L21} were used to calculate the distances and absolute magnitudes shown in Tables 1 and 2. Paper I, which used the DR2 Gaia parallaxes, showed that the majority of CVs from ZTF had absolute magnitudes at quiescence between 10$-$12, near the faint end of previous results in the literature \citep{W87,W95}. Figure 5 (right panel) shows a similar distribution for the second year, with an even larger number of systems at the fainter magnitudes of 12$-$13. This is likely due to the greater number of parallaxes of fainter objects available with Gaia EDR3. This increase also makes the distribution of parallax between known and candidate systems more equal (left panel of Figure 5). However, the 6 faintest absolute magnitudes (those $\ge$13.0) all have relatively close distances of 108-365 pc, meaning that they are intrinsically faint. The trends of increasing outburst amplitude and decreasing outburst frequency for the faintest absolute magnitudes (Figure 6) as seen in Paper 1 are also apparent, consistent with low mass transfer rates and low disk viscosity in these systems \citep{HSC95}, although the large scatter indicates there is not a simple relationship between these quantities. 

\section{Peculiar Light Curves}

There are several systems that have light curves that do not look like normal outbursts of dwarf novae. Included among these are the systems with high and low states. Two are shown in Figure 1 (ZTF2134-02 and ZTF2131+49), and three others are ZTF0434+03, ZTF2119+41 and ZTF2239+23, shown in Figure 7. Among these 5, ZTF2134-02, ZTF2119+41 are named as X-ray sources and ZTF0434+03 is listed as a possible (but not confirmed) quasar by \citet{DS11}. A redshift measurement can clarify the nature of this source. As noted above, the spectrum of ZTF2131+49 shows high excitation consistent with a magnetic white dwarf, while ZTF2134-02 looks more like a typical dwarf nova. Spectra of the other 3 can determine their correct classification. 

ZTF1736+75 and ZTF1756+02 (Figure 7) have features that show low amplitude outbursts, and a slight plateau or standstill at about one magnitude below their outburst magnitude. These are signatures that could classify them as Z Cam type systems, with relatively long orbital periods and high accretion rates that keep the disks near the limit for dwarf nova outbursts \citep{W95}. Spectra can determine the orbital period and reveal expected deep absorption lines with emission cores during the standstill states. Lastly, ZTF1848+41 has a large dip in the middle of its SOB which is quite distinctive, and indicative of a small class of dwarf novae with an extreme mass ratio due to a degenerate secondary \citep{K15}. Further monitoring will show if this is a peculiarity that is present after each SOB or if it is a unique occurrence.

\section{Completeness}
The GROWTH Marshal uses difference images in the alerts each night to produce candidates that are available to view for 5 nights at a time. Due to bad weather or instrument problems at the time of the rise to brightness, objects can be missed and not saved. Thus, this approach produces candidates and known CVs but is not complete. Recently, a machine learning (ML) method \citep{C20,vR21} to find various types of outbursting stars and variables has been accomplished using the entire existing light curves from the project. It generally finds more candidates than the Marshal alert method, but also
has flaws based on matching the light curves to correct object types and requiring an actual measurement at quiescence rather than an upper limit.
The ML method generally has a true positive rate for CVs of $\sim$25\% i.e. only 1 in 4 objects is an actual CV, as determined by visual inspection of the light curves and period determination to identify any periodic variables. Most of the false positives are irregular variables or 'bogus' light curves. To test the differences in the two methods, we compared the results found from the Marshal for the specific month of 2019 September with those objects from the ML set that had data obtained within that month. This comparison showed that the ML method found 227 objects while the GROWTH method found 55. The overlap between the two methods points out the differences. Including the past discoveries, the GROWTH Marshal found 78 of the 227 machine learning objects, thus missing 66\%, while the machine learning missed 36 (65\%) of the 55 found from the Marshal. Since there are no obvious intrinsic differences among the overlapping and missing groups of objects, the missed objects are likely due to the limitations in each method listed above.

Current efforts are underway to refine the ML capabilities by adding the misclassifications to the machine learning training set to 'teach' the algorithm how to better distinguish CVs from the false positives. At the current time, a new Fritz Marshal is in development as an alert broker for internal use by the ZTF collaboration and has replaced the GROWTH Marshal. It is expected to provide more flexibility in trying different filters and in cross-matching with multiwavelength databases that will ultimately lead to better identification of CV candidates. 

Another approach to finding CVs by using a periodicity search \citep{O20}, found about 60 new dwarf nova candidates in the DR1 database. Of these, 32 overlap with the GROWTH Marshal results.

\section{Conclusions}

The second year of ZTF alert filtering with the GROWTH Marshal has produced a list of 372 known and candidate CVs. Gaia parallaxes are available for almost half of these systems, and the resulting absolute magnitudes continue to show that most of the new systems being discovered are at the faint end of the distribution. The faintest ones are relatively closeby, and therefore likely have the lowest mass-transfer rates.  Several systems merit followup observations to provide their detailed classification. ZTF2131+49 shows high excitation \ion{He}{2} emission and could harbor a magnetic white dwarf. ZTF1736+75 and ZTF1756+02 have features resembling standstills in their light curves and may be Z Cam type systems with high mass-transfer rates. ZTF1848+41 had a peculiar SOB with a large decrease in brightness that appeared to divide the SOB in two making it a good candidate for having a degenerate companion. Ongoing machine learning methods appear to find a greater number of CVs using the entire data from the onset of the survey rather than from nightly alerts. Since each method produces different results, further refinements are needed and ongoing to obtain the optimum candidates for all types of CVs from ZTF.

\acknowledgments

PS,BK,CL and BD acknowledge funding from NSF grant AST-1514737.
A.Y.Q.H. is supported by a National Science Foundation Graduate Research Fellowship under Grant No.\,DGE-1144469. MC is supported by the David and Ellen Lee Prize Postdoctoral Fellowship at the California Institute of Technology. MLG acknowledges support from the DIRAC Institute in the Department of Astronomy at the University of Washington. The DIRAC Institute is supported through generous gifts from the Charles and Lisa Simonyi Fund for Arts and Sciences, and the Washington Research Foundation. 
This work was supported by the GROWTH project funded by the National Science Foundation under PIRE Grant No.\,1545949 and based on observations obtained with the Samuel Oschin Telescope 48-inch and the 60-inch Telescope at the Palomar Observatory as part of the Zwicky Transient Facility project. ZTF is supported by the NSF under grant AST-1440341 and a collaboration including Caltech, IPAC, the Weizmann Institute for Science,the Oskar Klein Center at Stockholm University, the University of Maryland, the University of Washington, Deutsches Elektronen-Synchrotron and Humboldt University, Los Alamos National Laboratories, the TANGO Consortium of Taiwan, the University of Wisconsin at Milwaukee, and Lawrence Berkeley national Laboratories. Operations are conducted by COO, IPAC and UW.
This work also makes use of observations from the Las Cumbres Observatory global telescope network.

\vspace{5mm}
\facilities{Keck:I;PO:1.2m}

\clearpage 
\startlongtable 
\begin{longrotatetable} 
\begin{deluxetable}{lccccccccccccl} 
\tabletypesize{\footnotesize} 
\tablewidth{0pt} 
\tablecolumns{14} 
\tablecaption{Known Confirmed CVs} 
\tablehead{
\colhead{ZTF} & \colhead{RA} & \colhead{Dec} & \colhead{b$^{\circ}$} & \colhead{$\Delta$mag} & \colhead{p (mas)} & \colhead{d (pc)} & \colhead{M} & \colhead{Out} & \colhead{Days} & \colhead{SDSS} & \colhead{CRTS} & \colhead{Spec\tablenotemark{a}} & \colhead{Other Surveys\tablenotemark{b}} }
\startdata 
18acahrug & 00:01:30.46 & +05:06:23.5 & -55.6 & 14.9-19.6 & --- & --- & --- & 3 & 233 & Y & Y & SD & AT2020qvd, 2M, G \\ 
18abspogs & 00:25:00.19 & +07:33:49.2 & -54.7 & 16.1-19.7 & $0.63\pm0.2$ & 1581 & 8.7 & 2 & 251 & Y & Y & SD & AT2019mvd, G \\ 
17aaakpyj & 00:35:35.71 & +46:23:52.2 & -16.4 & 14.4-19.4 & $2.1\pm0.04$ & 475 & 11.0 & 4 & 360 & --- & --- & --- & 2M, G, Gx \\ 
18abumkca & 00:36:40.28 & +23:08:31.3 & -39.6 & 15.5-20.9 & $2.01\pm0.17$ & 498 & 12.4 & 1 & 254 & Y & Y & SD & ASASSN-14dr, G \\ 
18abnzljg & 00:51:52.88 & +20:40:17.3 & -42.2 & 15.3-20.7 & $0.48\pm0.14$ & 2096 & 9.1 & 2 & 249 & Y & --- & --- & 2M, G, Gx \\ 
17aaaeoeh & 00:52:18.00 & +53:51:50.0 & -9.0 & 15.3-19.9 & $1.3\pm0.13$ & 772 & 10.5 & 6 & 344 & --- & --- & --- & V452 Cas, 2M, G \\ 
18actytbf & 01:11:57.61 & +35:17:24.3 & -27.4 & 13.7-20.0 & $1.52\pm0.23$ & 659 & 10.9 & 1 & 42 & Y & Y & --- & FN And, AT2019hst, 2M, G, Gx \\ 
18adcbymz & 01:13:06.73 & +21:52:50.2 & -40.7 & 14.3-20.7 & --- & --- & --- & SOB & 57 & Y & Y & --- & GV Psc, G, Gx \\ 
18aabfcyi & 01:15:32.20 & +37:37:35.5 & -25.0 & 14.0-19.3 & $1.67\pm0.08$ & 600 & 10.4 & 8 & 359 & --- & --- & --- & FO And, AT2020nyp, 2M, G, Gx \\ 
18abshhtu & 01:16:13.79 & +09:22:16.1 & -53.0 & 16.1-20.8 & $1.02\pm0.22$ & 982 & 10.8 & 3 & 253 & Y & Y & --- & G, Gx \\ 
18abuocqk & 01:25:39.37 & +32:23:08.1 & -29.9 & 12.2-19.8 & $3.84\pm0.07$ & 260 & 12.7 & 4 & 185 & Y & Y & --- & TY Psc, 2M, G, Gx \\ 
18abmjmhx & 01:36:37.02 & +32:00:40.1 & -29.9 & 13.6-18.4 & $1.43\pm0.05$ & 699 & 9.2 & 13 & 365 & Y & Y & --- & TW Tri, AT2018glm, 2M, G \\ 
18abscjio & 01:43:04.67 & +26:38:33.2 & -34.8 & 15.1-20.2 & $1.52\pm0.09$ & 657 & 11.1 & 6 & 256 & Y & Y & --- & AT2019nlz, G, Gx \\ 
17aaaswvx & 01:50:51.53 & +33:26:21.7 & -27.8 & 14.6-20.3 & $1.35\pm0.11$ & 739 & 11.0 & 3 & 268 & Y & --- & --- & G, Gx \\ 
17aaaewmi & 02:05:00.39 & +46:05:38.0 & -14.9 & 15.3-20.2 & $1.0\pm0.11$ & 1005 & 10.2 & 5 & 278 & Y & --- & --- & ASASSN-14gl, AT2018gcv, G \\ 
19adbgfwx & 02:15:30.03 & +57:17:54.0 & -3.7 & 14.1-20.8 & --- & --- & --- & SOB & 94 & --- & --- & --- & ASASSN-19ado, AT2019xim \\ 
18abomvne & 02:17:13.92 & +40:41:29.6 & -19.3 & 14.6-20.3 & --- & --- & --- & 1 & 86 & --- & Y & --- & KV And, AT2019kzl, G \\ 
18abtgioa & 02:25:00.47 & +32:59:55.5 & -25.9 & 15.7-20.3 & $1.97\pm0.11$ & 507 & 11.8 & 2 & 288 & --- & --- & --- & WY Tri, 2M, G, Gx \\ 
19abydbvw & 02:33:22.61 & +00:50:59.3 & -52.8 & 12.7-20.1 & --- & --- & --- & SOB & 184 & Y & Y & SD & HP Cet, G \\ 
18acdyfat & 02:42:16.15 & +35:40:47.2 & -22.0 & 15.3-20.5 & --- & --- & --- & 2 & 172 & --- & Y & --- & PU Per, 2M, G \\ 
18abnolsl & 02:42:53.40 & +38:04:03.2 & -19.8 & 15.2-20.2 & $1.17\pm0.22$ & 852 & 10.5 & 8 & 260 & --- & Y & --- & PV Per, AT2020yei, 2M, G, Gx \\ 
18abnzufs & 02:46:02.38 & +34:55:08.0 & -22.3 & 14.7-20.3 & $1.45\pm0.14$ & 691 & 11.1 & 8 & 248 & --- & --- & --- & V872 Per, AT2019rlf, 2M, G \\ 
18acebhoz & 02:56:39.66 & +37:08:23.2 & -19.3 & 18.2-20.9 & $1.46\pm0.39$ & 687 & 11.7 & 5 & 232 & Y & Y & --- & V372 Per, AT2018equ, G \\ 
19abxvmrz & 03:15:36.85 & +42:28:14.1 & -12.9 & 14.8-20.3 & --- & --- & --- & SOB,1 & 100 & Y & --- & --- & QY Per,  AT2019qxv, 2M, G \\ 
18abuxyhz & 03:34:49.87 & -07:10:47.9 & -46.5 & 14.4-20.3 & $1.57\pm0.17$ & 635 & 11.3 & 6 & 200 & Y & --- & --- & AT2018kwh, G, Gx \\ 
18abuylaz & 03:45:15.42 & -01:52:16.3 & -41.3 & 14.9-20.1 & $2.55\pm0.21$ & 392 & 12.1 & 3 & 220 & --- & --- & --- & G \\ 
18acrcryc & 03:51:56.99 & +25:25:27.8 & -21.8 & 15.7-19.8 & --- & --- & --- & 1 & 40 & --- & --- & --- & V1212 Tau \\ 
18aaaatwt & 04:02:39.07 & +42:50:45.6 & -7.4 & 12.4-18.9 & $3.96\pm0.06$ & 253 & 11.9 & 4 & 288 & --- & --- & --- & V1024 Per, G, Gx \\ 
18acpvfqh & 04:12:36.61 & +69:29:07.0 & 13.2 & 13.0-20.1 & $2.6\pm0.08$ & 385 & 12.2 & 2 & 223 & --- & --- & --- & NN Cam, 2M, G, Gx \\ 
17aaagyuc & 05:23:51.77 & +01:00:30.6 & -18.9 & 14.4-17.7 & $1.44\pm0.05$ & 695 & 8.5 & 2 & 228 & Y & --- & --- & BI Ori, AT2019cvt, 2M, G \\ 
17aabxrtj & 05:28:32.74 & +28:38:36.3 & -3.3 & 15.2-19.4 & --- & --- & --- & 4 & 267 & Y & --- & --- & 2M \\ 
17aadmkrk & 06:24:02.64 & +27:04:10.2 & 6.5 & 15.1-19.7 & $0.66\pm0.11$ & 1525 & 8.8 & 3 & 182 & Y & --- & --- & 2M, G \\ 
17aaavwmz & 06:36:54.60 & +00:02:17.2 & -3.2 & 12.3-19.2 & $3.04\pm0.05$ & 329 & 11.6 & 3 & 209 & Y & --- & --- & CW Mon, 2M, G \\ 
18abvtydv & 06:38:44.16 & +18:16:11.4 & 5.5 & 14.7-21.0 & $1.5\pm0.2$ & 667 & 11.9 & 3 & 181 & --- & --- & --- & UV Gem, G \\ 
17aabwjgy & 06:45:21.22 & +19:04:47.2 & 7.3 & 15.6-19.7 & --- & --- & --- & SOB,2 & 234 & --- & --- & --- & KT Gem, 2M, G \\ 
18aaawjmk & 07:19:12.13 & +48:58:34.7 & 24.5 & 15.6-20.6 & $1.19\pm0.23$ & 844 & 11.0 & 3 & 360 & --- & --- & --- & AT2020lhq, G \\ 
19acdufau & 07:32:08.12 & +41:30:08.7 & 24.8 & 16.6-18.4 & --- & --- & --- & 2 & 143 & Y & Y & SD & G \\ 
18aaeaftr & 07:45:31.89 & +45:38:29.0 & -36.5 & 12.7-19.9 & $3.22\pm0.22$ & 310 & 12.4 & SOB & 259 & Y & --- & SD & EQ Lyn, G \\ 
17aabxzol & 07:54:14.48 & +31:32:15.8 & 26.3 & 14.2-20.8 & $1.86\pm0.13$ & 537 & 12.2 & 1 & 65 & Y & Y & --- & G \\ 
17aabhbir & 07:58:53.01 & +16:16:45.1 & 22.1 & 15.0-19.0 & $4.88\pm0.04$ & 205 & 12.4 & 2 & 242 & Y & --- & SD & DW Cnc, 2M, G \\ 
17aaaipza & 08:03:03.89 & +25:16:26.9 & 26.2 & 15.3-18.8 & $1.0\pm0.21$ & 1001 & 8.8 & SOB,1 & 360 & Y & --- & SD & AT2019wcn, G \\ 
18aacktzu & 08:16:10.84 & +45:30:10.1 & 33.4 & 15.7-20.8 & --- & --- & --- & 1 & 255 & Y & Y & SD & AT2020dkd, G \\ 
18aakabcf & 08:20:19.41 & +47:47:31.0 & 34.3 & 17.5-19.9 & --- & --- & --- & 2 & 119 & Y & Y & SD & G, Gx \\ 
18accnopc & 08:36:42.68 & +53:28:38.1 & 37.0 & 12.9-19.7 & $6.2\pm0.06$ & 161 & 13.7 & 3 & 264 & Y & --- & SD & SW Uma, 2M, G, Gx \\ 
18acpoocp & 09:01:03.96 & +48:09:10.8 & 41.1 & 15.6-20.5 & $1.63\pm0.47$ & 615 & 11.6 & 3 & 343 & Y & Y & SD & G \\ 
18aaagsjn & 09:12:16.23 & +50:53:53.7 & 42.6 & 15.0-18.8 & $1.44\pm0.07$ & 694 & 9.6 & 9 & 362 & Y & Y & SD & DI Uma, 2M, G \\ 
18aalldmu & 09:19:35.66 & +50:28:25.0 & 43.9 & 15.6-20.7 & $0.63\pm0.2$ & 1597 & 9.7 & 3 & 362 & Y & Y & --- & G \\ 
18aabjgdh & 09:32:49.57 & +47:25:22.9 & 46.5 & 16.5-20.0 & $1.46\pm0.27$ & 686 & 10.8 & 1 & 349 & Y & --- & SD & G \\ 
18aalurns & 09:45:50.99 & -19:44:02.0 & 25.1 & 13.4-19.4 & $5.51\pm0.07$ & 181 & 13.1 & 1 & 198 & --- & Y & --- & NSV4618, AT2020aio, G \\ 
17aabhicw & 09:46:34.47 & +13:50:57.8 & 45.0 & 16.3-19.7 & $0.75\pm0.07$ & 1329 & 9.1 & 2 & 237 & Y & Y & SD & HY Leo, 2M, G, Gx \\ 
18aaaocpc & 09:46:36.55 & +44:46:44.7 & 49.3 & 14.6-20.8 & $2.59\pm0.15$ & 387 & 12.9 & 3 & 355 & Y & Y & SD & DV UMa, G, 2M \\ 
20aaeuecr & 09:47:59.82 & +06:10:44.1 & 41.7 & 13.5-20.0 & --- & --- & --- & SOB,2 & 80 & Y & Y & --- & G, MOT \\ 
18aclaxey & 10:15:39.81 & +73:26:04.8 & 39.4 & 14.7-20.2 & $1.41\pm0.27$ & 708 & 10.9 & 3 & 327 & --- & --- & --- & CP Dra, AT2018rv, G \\ 
18aaaujae & 10:23:20.28 & +44:05:09.4 & 55.9 & 14.7-20.3 & $1.56\pm0.15$ & 639 & 11.3 & 5 & 364 & Y & Y & SD & NSV 4838, PB 195, G, Gx \\ 
19aavnxej & 10:24:02.70 & +48:08:51.0 & 54.6 & 12.9-19.3 & $2.8\pm0.25$ & 357 & 11.5 & SOB & 31 & Y & Y & SD & G \\ 
18aabkmsj & 10:43:25.08 & +56:32:57.9 & 52.8 & 16.7-20.0 & --- & --- & --- & 1 & 48 & Y & Y & --- & G, Gx \\ 
18acvweuk & 10:52:15.27 & -06:43:26.4 & 45.5 & 15.8-20.1 & --- & --- & --- & 3 & 214 & --- & Y & --- & G, Gx \\ 
18aabszen & 11:00:14.72 & +13:15:51.8 & 60.6 & 14.3-20.0 & $1.81\pm0.14$ & 554 & 11.3 & 4 & 358 & Y & Y & SD & G \\ 
19aaaolka & 11:17:59.68 & +76:51:30.4 & 39.0 & 15.1-17.3 & --- & --- & --- & 2 & 182 & Y & --- & --- & G, Gx \\ 
18abcicny & 11:20:03.37 & +66:36:32.3 & 48.1 & 15.5-19.5 & --- & --- & --- & SOB,1 & 362 & Y & Y & SD & AT2016bkt, G, Gx \\ 
18acyerom & 12:40:58.03 & -01:59:19.3 & 60.8 & 13.7-19.4 & $1.52\pm0.47$ & 658 & 10.3 & SOB,3 & 361 & Y & Y & SD & AT2020njd, G \\ 
17aabxrbe & 12:44:26.26 & +61:35:14.4 & 55.5 & 15.1-20.6 & $1.45\pm0.1$ & 690 & 11.4 & 3 & 361 & Y & Y & SD & V351 UMa, 2M, G, Gx \\ 
17aabwtnr & 12:56:37.13 & +26:36:43.0 & 88.7 & 14.1-20.0 & $2.25\pm0.12$ & 444 & 11.8 & 3 & 365 & Y & Y & SD & GO Com, G \\ 
18aakpzqg & 13:43:23.14 & +15:09:16.9 & 73.0 & 13.5-20.8 & $2.29\pm0.14$ & 436 & 12.6 & 7 & 365 & Y & Y & SD & HW Boo, AT2020cmp, G, Gx \\ 
18abaloet & 15:28:57.86 & +03:49:11.5 & 45.8 & 16.6-18.3 & --- & --- & --- & 1 & 176 & Y & Y & SD & AT2020nrj, G \\ 
18aauylvz & 16:25:20.29 & +12:03:08.5 & 37.8 & 13.0-19.1 & $2.28\pm0.17$ & 438 & 10.9 & 3 & 366 & Y & Y & SD & 2M, G, Gx \\ 
18aaixpsq & 16:28:30.88 & +24:02:59.2 & 41.3 & 15.5-20.8 & $1.54\pm0.2$ & 651 & 11.7 & 3 & 365 & Y & Y & SD & AT2020gkw, G \\ 
18aaylslx & 16:42:48.51 & +13:47:51.4 & 34.7 & 15.3-20.2 & $1.89\pm0.09$ & 529 & 11.6 & 4 & 356 & Y & Y & SD & G \\ 
18abetdej & 17:06:09.67 & +14:34:51.6 & 30.0 & 14.0-20.9 & $1.97\pm0.12$ & 508 & 12.4 & 2 & 367 & Y & Y & --- & G \\ 
18ablxbjf & 18:08:35.85 & +10:10:29.6 & 14.1 & 14.7-18.9 & $3.41\pm0.06$ & 293 & 11.6 & 11 & 367 & --- & --- & --- & RXJ1808+10, 2M, G \\ 
18absqynr & 18:11:24.87 & -14:55:34.6 & 1.8 & 12.8-18.0 & $3.21\pm0.04$ & 312 & 10.5 & 2 & 362 & --- & --- & --- & 2M, G \\ 
19acdulxu & 18:14:14.23 & +30:43:36.8 & 20.9 & 18.1-19.2 & --- & --- & --- & 2 & 221 & --- & Y & --- & V1010 Her, 2M, G \\ 
18abjdnrd & 18:52:41.39 & +26:45:31.3 & 11.6 & 13.4-16.6 & $2.1\pm0.03$ & 477 & 8.2 & 5 & 363 & --- & --- & --- & CY Lyr, 2M, G, Gx \\ 
18aapaldh & 18:56:08.15 & +45:37:39.8 & 18.3 & 15.4-18.3 & $0.99\pm0.02$ & 1007 & 8.3 & 13 & 363 & --- & Y & --- & KIC9202990, 2M, G \\ 
18aanvxoj & 18:58:32.08 & +51:48:57.3 & 20.0 & 16.1-19.2 & $1.24\pm0.03$ & 808 & 9.7 & 4 & 366 & --- & --- & --- & HS1857+5144, 2M, G, KIC \\ 
18abbmmuf & 19:10:14.01 & +29:06:13.7 & 9.1 & 14.7-19.5 & $1.19\pm0.09$ & 843 & 9.9 & 7 & 340 & --- & --- & --- & V419 Lyr, G \\ 
18abgrsxw & 19:10:59.41 & +28:56:38.8 & 8.9 & 15.1-20.3 & $1.03\pm0.11$ & 969 & 10.4 & 3 & 279 & --- & --- & --- & V584 Lyr, AT2020znm, 2M, G \\ 
18abnowur & 19:26:10.82 & -10:15:30.5 & -12.3 & 12.6-20.3 & $3.33\pm0.13$ & 301 & 12.9 & 2 & 359 & --- & --- & --- & DH Aql, AT2020abfx, 2M, G, Gx \\ 
18abacgon & 19:33:53.63 & +14:17:45.6 & -2.7 & 12.6-20.1 & $3.24\pm0.14$ & 308 & 12.7 & SOB,1 & 151 & --- & --- & --- & KX Aql, G \\ 
18abcfvqc & 19:40:16.16 & +46:32:47.9 & 11.6 & 16.4-21.0 & --- & --- & --- & 4 & 131 & --- & --- & --- & G \\ 
18abnpdqe & 19:51:31.11 & +10:57:21.7 & -8.0 & 15.3-19.1 & $0.85\pm0.08$ & 1181 & 8.7 & 4 & 363 & Y & --- & --- & V1047 Aql, AT2018jqn, G \\ 
19abraqpf & 20:14:18.06 & +51:43:23.9 & 9.4 & 18.1-19.8 & --- & --- & --- & SOB & 40 & --- & --- & --- & AT2019onf \\ 
18abgyoaa & 20:16:49.97 & +53:12:24.3 & 9.9 & 14.6-20.0 & $1.01\pm0.12$ & 989 & 10.0 & 5 & 367 & --- & --- & --- & V767 Cyg, 2M, G \\ 
18abjgdih & 20:25:22.89 & +15:45:57.2 & -12.6 & 15.2-19.6 & $0.67\pm0.1$ & 1484 & 8.7 & 6 & 366 & --- & --- & --- & EZ Del, AT2020ljo, 2M, G \\ 
18abuktcs & 20:31:09.57 & +16:23:08.3 & -13.4 & 16.1-19.2 & $0.7\pm0.06$ & 1426 & 8.4 & 9 & 366 & --- & --- & --- & IS Del, 2M, G \\ 
18aazmwvg & 20:36:55.47 & +14:03:09.3 & -15.8 & 13.7-20.3 & $1.99\pm0.13$ & 503 & 11.8 & 4 & 357 & Y & --- & --- & HO Del, 2M, G \\ 
18abasoef & 21:04:04.69 & +46:31:13.5 & -0.2 & 11.0-20.1 & $9.23\pm0.07$ & 108 & 15.0 & HL & 364 & --- & --- & --- & AT2019lxx, G \\ 
19aavqcpp & 21:36:04.22 & +40:26:19.4 & -8.6 & 14.6-19.5 & $1.94\pm0.06$ & 516 & 10.9 & 4 & 158 & --- & --- & --- & V632 Cyg, 2M, G \\ 
18acattce & 22:16:31.14 & +29:00:19.8 & -22.7 & 12.9-19.3 & $2.5\pm0.05$ & 400 & 11.3 & SOB,1 & 154 & Y & --- & --- & V513 Peg, 2M, G, Gx \\ 
18abesoyi & 22:39:58.27 & +23:18:36.7 & -30.4 & 16.8-20.8 & --- & --- & --- & 7 & 226 & Y & Y & --- & ATel2654 \\ 
17aabulaf & 23:23:08.45 & +18:24:58.5 & -39.7 & 12.5-17.5 & $7.09\pm0.04$ & 141 & 11.8 & 3 & 341 & Y & --- & --- & IP Peg, 2M, G, Gx \\ 
17aabuphg & 23:27:02.13 & +50:07:12.7 & -10.5 & 15.8-18.9 & $0.73\pm0.05$ & 1377 & 8.2 & 5 & 276 & Y & --- & --- & BV And, 2M, G \\ 
18abwiccd & 23:40:20.66 & +76:42:10.3 & 14.4 & 16.9-20.6 & $1.79\pm0.05$ & 559 & 11.9 & 6 & 367 & --- & --- & --- & G \\ 
\enddata 
\tablenotetext{a}{SD=SDSS} 
\tablenotetext{b}{MOT=MASTEROT, G=Gaia, Gx=GALEX, 2M=2MASS} 
\end{deluxetable} 
\end{longrotatetable} 

\clearpage
\startlongtable
\begin{longrotatetable}
\begin{deluxetable}{lccccccccccccl}
\tabletypesize{\footnotesize}
\tablewidth{0pt}
\tablecolumns{14}
\tablecaption{CV Candidates}
\tablehead{
\colhead{ZTF} & \colhead{RA} & \colhead{Dec} & \colhead{b$^{\circ}$} & \colhead{$\Delta$mag} & \colhead{p(mas)} & \colhead{d(pc)} & \colhead{M} & \colhead{Out} & \colhead{Days} &  \colhead{SDSS} & \colhead{CRTS} & \colhead{Spec\tablenotemark{a}} &  \colhead{Other Surveys\tablenotemark{b}} }
\startdata
18abmnxli & 00:02:22.39 & +42:42:13.2 & -19.3 & 14.7-20.5 & $2.49\pm0.13$ & 402 & 12.5 & 6 & 353 & --- & Y & --- & ASASSN-13cx, G \\ 
19abpfdji & 00:06:58.81 & +03:51:04.1 & -57.2 & 17.1-21.0 & --- & --- & --- & 1 & 29 & --- & Y & --- & --- \\ 
18achnnpz & 00:08:20.31  & +77:31:18.9  & 14.8 & 17.1-19.2 & --- & --- & --- & 4 & 247 & --- & --- & --- & G, MOT  \\ 
18abmnxnv & 00:10:19.31 & +41:04:54.6 & -21.1 & 15.6-20.6 & --- & --- & --- & 3 & 353 & --- & Y & --- & G \\ 
18ablmdrf & 00:32:03.62  & +31:45:10.3  & -30.9 & 15.0-19.9 & $1.07\pm0.21$ & 935 & 10.0 & 1 & 351 & Y & --- & --- & G \\ 
18acrsqpx & 00:40:09.09 & +54:03:08.2 & -8.8 & 17.2-19.2 & --- & --- & --- & 1 & 33 & --- & --- & --- & --- \\ 
18abtyikp & 00:45:27.56 & +50:32:15.1 & -12.3 & 13.7-20.3 & $2.66\pm0.22$ & 376 & 12.4 & SOB & 231 & --- & --- & --- & G, Gx, MOT \\ 
19acgerbh & 00:45:41.95 & +47:01:24.1 & -15.8 & 17.1-18.7 & --- & --- & --- & SOB & 32 & --- & --- & --- & G \\ 
18abwwckd & 00:50:20.60 & +33:19:17.6 & -29.6 & 16.7-20.5 & --- & --- & --- & SOB & 119 & Y & Y & --- & G \\ 
18acfhcfx & 00:57:18.50  & +67:54:05.2  & 5.0 & 14.8-17.1 & $1.12\pm0.01$ & 893 & 7.3 & 2 & 95 & --- & --- & --- & 2M, G  \\ 
19acyfeuy & 00:59:09.74  & +34:38:35.6  & -28.2 & 12.9-19.9 & --- & --- & --- & SOB,1  & 88 & Y & Y & --- & AT2019wnm, G, Gx  \\ 
18absondn & 01:00:57.45  & +57:14:35.3  & -5.6 & 16.6-20.9 & $1.52\pm0.41$ & 657 & 11.8 & 2 & 175 & --- & --- & --- & AT2019dpn, G \\ 
18acquwya & 01:03:28.92 & +33:18:20.4 & -29.5 & 17.2-20.9 & --- & --- & --- & 3 & 113 & Y & Y & --- & G \\ 
18acakbbq & 01:09:25.86 & +52:10:21.3 & -10.6 & 17.5-20.7 & --- & --- & --- & 7 & 353 & --- & --- & --- & G, MOT \\ 
18abckswk & 01:09:46.94 & +68:00:34.2 & 5.2 & 17.0-20.0 & $0.94\pm0.1$ & 1060 & 9.9 & 12 & 356 & --- & --- & --- & G \\ 
19abqgkdf & 01:16:12.52 & +32:38:39.4 & -29.9 & 17.4-20.4 & --- & --- & --- & SOB & 123 & Y & Y & --- & G \\ 
19acasute & 01:18:54.01 & +38:02:32.1 & -24.5 & 16.7-19.7 & --- & --- & --- & 1 & 31 & --- & --- & --- & G \\ 
18abmexrv & 01:20:29.83 & +58:00:19.5 & -4.7 & 17.7-20.0 & --- & --- & --- & 9 & 359 & --- & --- & --- & G \\ 
20aafdhqm & 01:29:49.25  & +51:53:18.2  & -10.5 & 16.7-20.0 & --- & --- & --- & SOB & 42 & Y & --- & --- & AT2020xg \\ 
19abochaj & 01:32:02.78 & -10:43:57.8 & -71.0 & 15.9-20.4 & --- & --- & --- & 4 & 51 & Y & Y & --- & ASASSN-14kk, G \\ 
18acrsmyu & 01:33:07.56  & +41:07:18.7  & -21.1 & 15.3-20.1 & --- & --- & --- & 2 & 244 & Y & Y & --- & AT2019lfs  \\ 
19abylhhp & 01:57:46.15 & +51:10:23.9 & -10.3 & 16.1-18.8 & $1.23\pm0.17$ & 814 & 9.2 & 1 & 31 & --- & --- & --- & ASASSN-15aw, G, MOT \\ 
18abycivz & 02:10:27.53 & +50:21:25.9 & -10.6 & 17.2-19.6 & --- & --- & --- & 3 & 229 & --- & --- & --- & AT2018gun, G \\ 
18abtrvrd & 02:13:17.19  & +46:06:43.4  & -14.4 & 15.3-20.7 & --- & --- & --- & 8 & 269 & Y & --- & --- & G, Gx \\ 
18abslfyk & 02:24:11.44 & +35:59:18.4 & -23.2 & 15.6-20.6 & --- & --- & --- & 3 & 211 & --- & Y & --- & G \\ 
18acrcorr & 02:25:20.76 & +25:31:22.5 & -32.7 & 17.8-20.2 & --- & --- & --- & 3 & 117 & Y & Y & --- & G \\ 
17aabulav & 02:32:11.61  & +30:36:34.4  & -27.4 & 16.6-19.8 & --- & --- & --- & 3 & 290 & Y & --- & --- & G \\ 
19abzswmj & 02:34:06.06 & +38:41:42.4 & -19.9 & 16.2-18.4 & --- & --- & --- & SOB & 96 & --- & --- & --- & MOT \\ 
19aceluqi & 02:35:03.96 & +32:40:37.2 & -25.3 & 17.4-19.8 & --- & --- & --- & SOB & 167 & --- & Y & --- & AT2020qvl, G \\ 
18acdxkxr & 02:35:08.09  & +79:41:55.8  & 17.8 & 18.2-20.5 & --- & --- & --- & 7 & 248 & Y & --- & --- & G \\ 
18abrpupq & 02:36:38.02  & +11:11:56.4  & -44.0 & 15.4-20.6 & $0.89\pm0.27$ & 1123 & 10.3 & 4 & 206 & --- & --- & --- & G \\ 
18acslhet & 02:47:12.78  & +20:10:42.1  & -35.0 & 18.0-20.8 & --- & --- & --- & 2 & 131 & Y & Y & --- & Gx  \\ 
18acsytgy & 02:48:50.27  & +40:14:48.2  & -17.3 & 16.0-17.7 & --- & --- & --- & SOB,2 & 196 & --- & Y & --- & MOT \\ 
18abtteya & 02:56:12.87 & -10:33:59.4 & -56.3 & 13.6-20.4 & $2.08\pm0.33$ & 480 & 12.0 & SOB & 119 & --- & Y & --- & AT2019qxu, G \\ 
19abudvoz & 03:06:45.06 & +46:09:12.9 & -10.6 & 18.1-19.1 & --- & --- & --- & 2 & 31 & --- & --- & --- & AT2016ayj \\ 
18acuirtm & 03:20:33.21 & +06:43:05.1 & -40.5 & 17.6-18.8 & $1.4\pm0.45$ & 713 & 9.6 & 3 & 145 & --- & Y & --- & G \\ 
17aadpgag & 03:26:27.26  & +07:07:44.4 & -39.2 & 17.3-19.9 & --- & --- & --- & 3 & 194 & --- & --- & --- & G \\ 
19abfvpfe & 03:42:06.37  & +32:37:07.6  & -17.8 & 18.2-19.2 & --- & --- & --- & 2 & 190 & --- & Y & --- & --- \\ 
18aaaatzj & 04:21:51.43  & +30:41:38.2  & -13.4 & 17.0-19.7 & --- & --- & --- & 9 & 249 & Y & --- & --- & AT2019nxm, G \\ 
18abupvkl & 04:24:34.16  & +00:14:18.6  & -32.0 & 15.7-20.6 & $1.08\pm0.14$ & 926 & 10.8 & 3 & 209 & Y & Y & --- & G \\ 
19aainjes & 04:34:44.49  & +03:06:15.9  & -28.3 & 18.4-20.4 & --- & --- & --- & HL & 200 & --- & Y & --- & --- \\ 
18adbmque & 04:49:58.11 & +31:28:02.4 & -8.4 & 17.1-19.7 & $0.99\pm0.15$ & 1007 & 9.6 & 3 & 89 & --- & --- & --- & G \\ 
19aamfurm & 04:59:55.87  & +77:11:18.1  & 20.6 & 15.6-19.5 & $1.32\pm0.35$ & 758 & 10.1 & 2 & 154 & --- & --- & --- & G, Gx  \\ 
18acbvwtm & 05:02:19.77  & +41:30:58.1  & -0.2 & 17.1-17.8 & --- & --- & --- & 1 & 119 & --- & --- & --- & --- \\ 
20aagqcbq & 05:09:12.70  & +49:09:00.7  & 5.4 & 15.3-19.8 & --- & --- & --- & SOB & 48 & --- & --- & --- & AT2020alt \\ 
19abytggf & 05:09:17.76 & +36:39:07.3 & -2.0 & 17.8-20.7 & --- & --- & --- & SOB & 101 & --- & --- & --- & AT2019qhx \\ 
19aaexvaq & 05:10:19.76 & +27:13:48.8 & -7.42 & 15.9-19.6 & --- & --- & --- & 5 & 205 & Y & --- & --- & --- \\ 
17aaceoyy & 05:10:54.56 & +18:10:42.8 & -12.5 & 16.2-20.3 & $2.52\pm0.44$ & 397 & 12.3 & 2 & 156 & --- & Y & --- & AT2020eus, G, MOT \\ 
18acamvve & 05:14:27.93  & +42:35:36.7  & 2.3 & 15.8-19.6 & --- & --- & --- & SOB & 38 & --- & --- & --- & 2M, G  \\ 
20aadfvvl & 05:15:30.00  & +03:22:12.8  & -19.5 & 17.1-17.1 & --- & --- & --- & 1 & 31 & --- & Y & --- & --- \\ 
17aabvsip & 05:25:21.94  & -10:50:15.3  & -24.0 & 16.4-18.2 & $0.76\pm0.25$ & 1322 & 7.6 & 3 & 139 & --- & Y & --- & AT2016dtl, G \\ 
19acdyyei & 05:34:44.74 & +23:10:45.2 & -5.1 & 17.3-19.3 & --- & --- & --- & SOB & 45 & --- & --- & --- & --- \\ 
18acyuvvq & 05:36:28.43  & +13:53:39.9  & -9.7 & 16.7-21.0 & --- & --- & --- & 2 & 142 & --- & Y & --- & G, MOT \\ 
17aacqwwi & 05:37:14.79  & +15:54:17.6  & -8.5 & 17.0-19.7 & $0.36\pm0.12$ & 2755 & 7.5 & 8 & 258 & --- & Y & --- & 2M, G, MOT \\ 
18abtzxrm & 05:48:02.03 & +39:44:02.2 & 6.0 & 18.1-20.0 & --- & --- & --- & 5 & 173 & --- & --- & --- & AT2016dyy, G \\ 
18aaaecnk & 05:48:44.38  & +49:00:34.8  & 10.7 & 16.6-20.6 & --- & --- & --- & SOB & 39 & --- & --- & --- & AT2018jl  \\ 
18aabicgg & 05:52:51.93  & +69:46:35.7  & 20.5 & 17.2-20.9 & --- & --- & --- & SOB,1 & 175 & --- & Y & --- & AT2016gcu, G \\ 
18aaaasra & 06:09:04.24  & +47:32:54.1  & 13.1 & 18.0-19.2 & --- & --- & --- & SOB,1 & 166 & --- & --- & --- & G \\ 
19accsokd & 06:10:54.00  & +07:09:18.6  & -5.7 & 18.3-19.0 & --- & --- & --- & 2 & 130 & --- & --- & --- & --- \\ 
18aaaasnn & 06:11:06.13 & +57:37:34.0 & 17.6 & 17.1-20.8 & --- & --- & --- & SOB & 163 & --- & --- & --- & G \\ 
17aaagnwn & 06:18:01.44  & +22:22:28.7  & 3.1 & 15.7-19.8 & --- & --- & --- & SOB & 58 & --- & --- & F & AT2019sgf, ASASSN-19yt, G \\ 
17aabvptl & 06:23:23.43 & +73:07:43.1 & 23.9 & 14.9-20.0 & $1.03\pm0.11$ & 967 & 10.1 & 6 & 294 & --- & --- & --- & AT2017avd, G, MOT \\ 
19acnlgks & 06:23:34.63  & -15:24:13.6  & -13.0 & 15.9-18.1 & --- & --- & --- & SOB & 36 & --- & Y & --- & AT2019uxa \\ 
17aabwdlv & 06:24:29.70 & +00:21:05.8 & -5.8 & 15.7-20.5 & $2.2\pm0.23$ & 455 & 12.2 & 4 & 219 & Y & --- & --- & AT2017gik, G \\ 
18achcvti & 06:33:19.90  & +22:09:30.5  & 6.1 & 17.1-20.6 & $1.82\pm0.51$ & 548 & 11.9 & 1 & 31 & --- & --- & --- & G \\ 
18aablttq & 06:34:37.96  & +41:20:46.4  & 14.6 & 16.1-19.9 & $1.49\pm0.3$ & 671 & 10.8 & 6 & 285 & --- & --- & --- & G, Gx \\ 
17aadiqwv & 06:34:39.49  & +32:25:18.7  & 10.9 & 16.9-21.0 & --- & --- & --- & 6 & 258 & --- & --- & --- & G \\ 
18acclkfa & 06:38:50.58  & +01:59:19.9  & -1.9 & 16.1-20.0 & $1.37\pm0.22$ & 728 & 10.7 & 2 & 190 & --- & --- & --- & AT2017haq, G \\ 
19aamfzhw & 06:39:17.63  & +57:23:10.4  & 21.0 & 15.7-18.6 & --- & --- & --- & 3 & 173 & --- & Y & --- & G, Gx  \\ 
18abzbukt & 06:42:33.28  & +40:02:33.2  & 15.5 & 16.7-18.2 & --- & --- & --- & SOB & 82 & --- & --- & --- & AT2018tl \\ 
18aabrxmh & 06:44:42.03  & +34:47:11.6  & 13.8 & 15.6-20.0 & $1.39\pm0.17$ & 719 & 10.7 & 4 & 268 & --- & --- & --- & AT2020rqq, G, Gx  \\ 
19abzrbxx & 06:46:08.21 & +40:33:04.9 & 16.3 & 14.8-20.5 & --- & --- & --- & SOB & 166 & --- & Y & --- & --- \\ 
19aajybch & 06:46:23.95  & +27:10:13.0  & 11.0 & 18.4-19.7 & --- & --- & --- & SOB & 37 & Y & --- & --- & AT2018bux \\ 
17aabshnt & 06:47:25.65 & +49:15:41.9 & 19.6 & 14.3-19.9 & $3.5\pm0.09$ & 286 & 12.6 & 4 & 280 & --- & --- & --- & G, Gx, MOT \\ 
17aabypst & 06:51:49.91  & +30:26:34.9  & 13.4 & 16.2-20.0 & --- & --- & --- & 6 & 277 & --- & --- & --- & 2M, G, Gx  \\ 
19acxffoy & 06:56:08.18  & +74:44:55.8  & 26.4 & 14.7-20.1 & --- & --- & --- & SOB & 57 & --- & --- & --- & AT2019wda, G, Gx, MOT  \\ 
18abyemxi & 06:56:20.46 & +35:56:18.9 & 16.5 & 15.7-20.6 & $2.39\pm0.46$ & 418 & 12.5 & 2 & 205 & --- & --- & --- & G \\ 
18aagqeeo & 06:58:56.51  & +50:49:34.5  & 21.8 & 17.3-19.1 & --- & --- & --- & 5 & 283 & --- & Y & --- & G, Gx, MOT  \\ 
18acusatr & 07:00:27.27 & +55:29:51.5 & 23.3 & 17.7-19.2 & --- & --- & --- & SOB & 232 & --- & --- & --- & G, MOT \\ 
19aamyzpg & 07:03:22.89  & +16:29:48.7  & 10.0 & 17.4-20.4 & --- & --- & --- & 2 & 149 & --- & --- & --- & --- \\ 
20aaoisnq & 07:13:11.20  & +30:15:31.6  & 17.6 & 14.1-16.6 & --- & --- & --- & SOB & 49 & Y & Y & --- & AT2020dnb \\ 
18aabrbjr & 07:20:07.38  & +45:16:11.4  & 23.7 & 16.5-20.5 & --- & --- & --- & 9 & 361 & --- & Y & --- & G, MOT \\ 
17aaasemy & 07:23:42.58 & +22:00:06.1 & 16.7 & 16.6-20.1 & --- & --- & --- & 5 & 181 & --- & Y & --- & AT2019bfs, G \\ 
18aaawuqj & 07:26:40.63  & +16:26:48.7  & 15.1 & 16.7-19.8 & $0.46\pm0.12$ & 2174 & 8.1 & 6 & 174 & Y & --- & --- & G, Gx  \\ 
19aarhrcv & 07:33:25.47  & +37:37:44.7  & 24.0 & 15.7-20.7 & --- & --- & --- & SOB & 48 & Y & Y & --- & AT2016arp, MOT \\ 
18adbgtis & 07:37:58.55  & +20:55:44.6  & 19.3 & 15.1-20.4 & --- & --- & --- & 1 & 30 & Y & Y & --- & G \\ 
18aaawmpv & 07:38:57.99  & +10:38:58.2  & 15.3 & 17.3-19.8 & --- & --- & --- & 3 & 144 & Y & Y & --- & G \\ 
18aaacggd & 07:44:00.47 & +41:55:03.5 & 27.1 & 16.6-20.0 & --- & --- & --- & 1 & 33 & Y & Y & --- & G \\ 
20aalztwm & 07:55:35.47  & -10:19:50.3  & 9.2 & 14.9-20.3 & --- & --- & --- & 1 & 39 & Y & --- & --- & AT2018ipf, G \\ 
17aabnzdk & 08:03:07.00  & +28:48:55.9  & 27.4 & 15.7-20.6 & --- & --- & --- & SOB,1  & 361 & Y & Y & --- & AT2020lqv, G, Gx \\ 
18achypzv & 08:22:12.35  & -01:21:46.4  & 19.4 & 15.8-19.4 & $0.81\pm0.04$ & 1235 & 8.9 & 2 & 227 & Y & Y & --- & 2M, G  \\ 
18acurpxm & 08:24:08.12  & +13:31:20.4  & 26.6 & 15.1-20.7 & --- & --- & --- & SOB,1  & 156 & Y & --- & --- & AT2018kwl \\ 
19aahvgyg & 09:08:52.20  & +07:16:39.2  & 33.8 & 18.3-19.9 & --- & --- & --- & 2 & 138 & Y & --- & --- & G, Gx \\ 
19aaaczho & 09:28:39.33  & +00:59:44.5  & 34.9 & 16.9-20.6 & --- & --- & --- & 3 & 207 & Y & Y & --- & --- \\ 
18acnnaav & 09:39:55.79  & +69:56:45.6  & 39.3 & 15.5-21.0 & $1.16\pm0.21$ & 861 & 11.3 & SOB,2 & 357 & --- & Y & --- & ASASSN-13ai, G, Gx \\ 
19aapcvao & 09:42:52.28 & -19:36:53.1 & 24.7 & 14.8-19.7 & $1.57\pm0.29$ & 636 & 10.7 & 1 & 4 & --- & Y & --- & AT2019dox, G \\ 
19aadnhaw & 10:11:43.78  & +57:18:13.8  & 48.8 & 16.0-20.9 & $1.17\pm0.37$ & 857 & 11.2 & SOB & 48 & Y & Y & --- & AT2020bvg, G, Gx \\ 
18aaadgen & 10:25:20.91  & +16:11:46.2  & 54.5 & 16.8-19.5 & --- & --- & --- & 3 & 278 & Y & Y & --- & G \\ 
20aakmtap & 11:23:32.03  & +43:17:17.6  & 66.0 & 15.4-20.5 & --- & --- & --- & SOB & 56 & Y & Y & --- & G \\ 
20aaiomlf & 11:29:19.16  & +37:34:07.9  & 69.8 & 16.7-20.5 & --- & --- & --- & SOB & 50 & Y & Y & --- & AT2020bxa \\ 
20aavxpoi & 12:40:57.32  & +12:11:13.3  & 74.9 & 18.9-21.0 & --- & --- & --- & 1 & 28 & --- & Y & --- & --- \\ 
18aabrqzd & 12:43:12.05  & +43:31:59.5  & 73.5 & 15.0-19.3 & --- & --- & --- & 3 & 353 & Y & Y & --- & ASASSN-13ao, G  \\ 
17aacnfia & 12:43:46.19  & +16:05:03.8  & 78.8 & 15.8-20.9 & $2.74\pm0.52$ & 365 & 13.1 & 1 & 149 & Y & Y & --- & G, MOT \\ 
18aabqind & 12:59:05.73  & +24:26:33.2  & 86.8 & 17.1-19.6 & --- & --- & --- & 4 & 199 & Y & Y & --- & G \\ 
19aavleeb & 13:19:37.93  & +09:58:28.7  & 71.6 & 18.3-21.0 & $6.22\pm1.63$ & 161 & 15.0 & 2 & 127 & Y & Y & --- & G \\ 
18acaxrth & 13:32:00.81  & -04:54:10.6  & 56.5 & 15.9-18.6 & $1.05\pm0.26$ & 952 & 8.7 & 2 & 305 & --- & Y & --- & G, Gx \\ 
19abahrdm & 14:05:13.28 & -06:18:17.4 & 52.1 & 16.2-20.1 & --- & --- & --- & 1 & 19 & --- & Y & --- & --- \\ 
18aaqlkdd & 14:35:50.02 & +59:21:34.3 & 53.1 & 14.9-20.6 & --- & --- & --- & 3 & 357 & Y & --- & --- & G, Gx, MOT \\ 
18abdricg & 14:59:21.84 & +35:48:05.8 & 61.5 & 15.9-21.0 & --- & --- & --- & 2 & 256 & --- & Y & --- & G \\ 
18aajpqbj & 15:11:09.80  & +57:41:00.2  & 50.8 & 17.7-20.2 & --- & --- & --- & 2 & 366 & Y & --- & --- & Gx \\ 
18aakticr & 15:13:32.97 & +70:37:22.5 & 42.1 & 16.1-20.7 & $2.02\pm0.26$ & 494 & 12.2 & 3 & 361 & --- & --- & --- & AT2017djy, G \\ 
18aayjsuk & 15:34:57.24 & +50:56:17.1 & 51.1 & 17.1-20.0 & --- & --- & --- & 3 & 302 & Y & --- & --- & AT2018cij, G \\ 
19aanoxtr & 15:36:27.17 & -16:09:25.8 & 31.1 & 15.3-20.2 & $1.24\pm0.22$ & 805 & 10.7 & 2 & 328 & --- & Y & --- & G \\ 
18aagwkwg & 15:55:40.22  & +36:46:43.0  & 50.1 & 16.5-21.0 & --- & --- & --- & 3 & 366 & Y & Y & --- & AT2019pux, G, Gx \\ 
18aathgvk & 16:23:23.31  & +78:26:03.8  & 33.7 & 13.2-20.5 & $2.7\pm0.09$ & 370 & 12.7 & 3 & 362 & Y & --- & --- & AT2018izu, G, Gx, MOT  \\ 
19aamkwxk & 16:27:16.74 & +04:06:02.9 & 33.7 & 13.5-20.7 & $4.21\pm0.24$ & 237 & 13.9 & SOB & 363 & --- & Y & --- & AT2019kwk, G \\ 
19aatmoqr & 16:39:08.92 & -17:35:53.7 & 19.0 & 17.1-20.6 & --- & --- & --- & 1 & 10 & --- & Y & --- & AT2019ekm \\ 
19aaqadty & 16:41:23.72 & -20:30:45.9 & 16.8 & 17.3-20.4 & $2.08\pm0.5$ & 481 & 12.0 & 2 & 351 & --- & Y & --- & AT2019jwu, G \\ 19aamwrrl & 16:41:26.29  & -19:37:28.1  & 17.3 & 17.3-20.0 & --- & --- & --- & 3 & 363 & --- & Y & --- & G \\ 
18abrwqal & 16:49:50.39 & +03:58:34.5 & 28.7 & 13.7-20.4 & $2.4\pm0.19$ & 417 & 12.3 & SOB,2 & 360 & Y & Y & --- & G \\ 
18abcubvf & 16:52:54.02 & +12:29:25.2 & 5.35 & 18.0-20.2 & --- & --- & --- & SOB,1 & 295 & --- & --- & --- & 2M \\ 
18absmovx & 16:55:38.86  & -04:09:22.5  & 23.4 & 18.4-20.4 & --- & --- & --- & 2 & 266 & Y & Y & --- & G \\ 
18abjydcz & 17:06:21.15  & -20:27:17.7  & 12.1 & 16.2-19.1 & --- & --- & --- & 2 & 363 & --- & --- & --- & 2M, G \\ 
18ablwpza & 17:08:52.08  & +01:26:55.6  & 23.4 & 14.2-20.3 & $1.6\pm0.1$ & 625 & 11.3 & 4 & 363 & --- & Y & --- & 2M, G \\ 
19aatmmgy & 17:21:48.83 & -05:17:13.3 & 17.2 & 15.9-20.4 & --- & --- & --- & 2 & 335 & --- & Y & --- & G  \\ 
18abbwhic & 17:22:35.82 & +18:05:11.6 & 27.5 & 14.3-19.9 & $2.09\pm0.12$ & 478 & 11.5 & SOB & 364 & Y & --- & --- & G \\ 
18aayzpbr & 17:29:51.59 & +22:08:08.0 & 27.3 & 14.8-18.5 & $0.91\pm0.09$ & 1105 & 8.3 & 8 & 366 & --- & Y & --- & G \\ 
19abmuxjv & 17:30:02.94 & +48:21:18.3 & 33.1 & 15.7-19.6 & --- & --- & --- & SOB,1 & 293 & --- & Y & --- & AT2018jaf, G \\ 
19abbtyln & 17:35:23.14 & -07:03:46.8 & 13.4 & 17.1-20.6 & --- & --- & --- & SOB & 38 & --- & Y & --- & AT2019kmg  \\ 
19aavtdeb & 17:35:28.26 & +36:31:06.6 & 30.3 & 17.1-20.1 & --- & --- & --- & 3 & 366 & --- & Y & --- & G \\ 
18abdggdv & 17:36:15.09  & +07:16:55.7  & 20.0 & 15.0-19.9 & --- & --- & --- & 3 & 365 & Y & --- & --- & ASASSN-14aj, G \\ 
18aaxuusk & 17:36:17.94 & +75:21:22.8 & 30.9 & 16.6-20.8 & --- & --- & --- & 2 & 362 & --- & --- & --- & G, MOT \\ 
19aaodsxt & 17:36:45.24  & +11:05:28.5  & 21.5 & 16.8-20.6 & $4.19\pm1.32$ & 239 & 13.7 & 2 & 286 & --- & --- & --- & G \\ 
18aakgpni & 17:38:50.99  & +29:23:11.2  & 27.7 & 17.0-21.0 & --- & --- & --- & SOB,4 & 366 & Y & Y & --- & AT2018hlb, G \\ 
18aakgoxi & 17:43:05.75  & +23:11:08.7  & 24.8 & 15.8-21.0 & --- & --- & --- & 7 & 366 & --- & Y & --- & AT2019pqu, G, MOT  \\ 
19abeamvv & 17:56:12.06 & -12:32:11.0 & 6.2 & 17.1-19.1 & --- & --- & --- & 1 & 32 & --- & --- & --- & --- \\ 
18abmbdyk & 17:56:22.38 & +02:58:04.1 & 13.6 & 17.0-19.1 & --- & --- & --- & 2 & 360 & --- & --- & --- & G \\ 
19aatmlxt & 17:57:43.80 & +19:10:50.2 & 20.2 & 15.8-19.9 & --- & --- & --- & SOB & 22 & --- & Y & --- & --- \\ 
18abnjsiq & 18:02:26.28 & +00:55:37.7 & 11.3 & 16.5-20.5 & --- & --- & --- & 2 & 152 & Y & --- & --- & AT2016bpr, G \\ 
18ablrnkx & 18:08:45.61 & +15:21:01.1 & 16.2 & 15.6-20.7 & $1.75\pm0.32$ & 571 & 12.0 & 8 & 349 & --- & --- & --- & AT2020rvb, G \\ 
19aazxpgk & 18:10:55.20 & +07:40:42.0 & 12.5 & 15.3-19.8 & --- & --- & --- & SOB & 26 & --- & --- & --- & ASASSN-15jw, G \\ 
18abucowm & 18:12:59.23  & +04:25:12.3  & 10.6 & 16.4-19.0 & --- & --- & --- & 5 & 319 & --- & --- & --- & G \\ 
18aarsonl & 18:13:29.60 & +45:36:16.6 & 25.4 & 17.5-20.2 & $0.65\pm0.03$ & 1539 & 9.3 & 3 & 364 & --- & Y & --- & G \\ 
18aapmalg & 18:14:06.57 & +39:05:55.1 & 23.6 & 17.8-21.0 & --- & --- & --- & 8 & 322 & --- & Y & --- & G \\ 
19aargcuk & 18:14:07.92 & +14:19:27.1 & 14.6 & 18.1-19.9 & --- & --- & --- & 3 & 366 & --- & --- & --- & AT2019gwe \\ 
19aayledy & 18:17:27.07 & +55:25:09.3 & 26.8 & 17.8-19.9 & --- & --- & --- & 3 & 310 & --- & --- & --- & --- \\ 
19abjdhzm & 18:18:09.45 & -16:37:26.5 & -0.4 & 15.7-19.4 & $1.36\pm0.21$ & 736 & 10.1 & 1 & 300 & --- & --- & --- & G \\ 
18abfwvvz & 18:20:51.21 & +49:12:00.5 & 25.0 & 15.3-20.8 & --- & --- & --- & SOB & 309 & --- & Y & --- & G \\ 
18abfsdut & 18:26:14.75 & +20:01:09.0 & 14.4 & 17.4-20.5 & --- & --- & --- & SOB,4 & 339 & --- & --- & --- & G \\ 
18ablwwzj & 18:32:00.54 & +19:09:37.6 & 12.8 & 16.5-20.5 & $1.79\pm0.52$ & 557 & 11.8 & 1 & 95 & --- & --- & --- & AT2017kai, G \\ 
18abcwxoh & 18:33:12.18 & +21:36:33.1 & 13.5 & 17.6-19.9 & --- & --- & --- & 5 & 366 & Y & --- & --- & AT2019sge, G \\ 
18aatluwz & 18:33:42.06 & +65:40:18.9 & 26.4 & 15.1-20.5 & $1.42\pm0.22$ & 702 & 11.2 & 3 & 268 & --- & Y & --- & AT2019kqk, G, MOT \\ 
18abshyar & 18:40:01.59 & +46:51:04.5 & 21.3 & 19.6-20.7 & --- & --- & --- & 1 & 33 & --- & Y & --- & --- \\ 
18abutmwk & 18:40:58.24 & +49:55:03.1 & 22.0 & 17.7-20.0 & --- & --- & --- & 1 & 34 & --- & Y & --- & G \\ 
18abdcofd & 18:46:59.47 & +12:04:24.2 & 6.5 & 17.6-20.3 & --- & --- & --- & 6 & 366 & --- & --- & --- & AT2019bkn, G \\ 
19accvbkv & 18:48:51.30 & +41:39:59.4 & 18.1 & 16.1-20.3 & --- & --- & --- & SOB & 54 & --- & --- & --- & AT2019sst \\ 
18abcubvf & 18:52:54.02 & +12:29:25.2 & 5.3 & 17.9-19.2 & --- & --- & --- & SOB,1 & 295 & --- & --- & --- & --- \\ 
18aarkacj & 18:53:09.61 & +59:45:07.2 & 23.0 & 14.9-20.3 & $1.56\pm0.09$ & 641 & 11.3 & 2 & 365 & --- & Y & --- & G \\ 
18abjbtnq & 18:53:30.61 & -01:28:16.3 & -1.1 & 13.0-19.4 & $4.53\pm0.2$ & 221 & 12.7 & 1 & 357 & Y & --- & --- & G \\ 
19abgbyqo & 18:54:19.33 & +31:05:20.3 & 13.1 & 18.7-20.3 & --- & --- & --- & 1 & 269 & --- & --- & --- & --- \\ 
18abfwzbg & 18:54:42.68 & +63:25:38.0 & 23.7 & 17.1-20.2 & --- & --- & --- & 7 & 329 & Y & --- & --- & G \\ 
19abbwmem & 18:57:20.56 & +36:22:40.4 & 14.6 & 15.4-20.8 & --- & --- & --- & 4 & 98 & --- & --- & --- & AT2019ijd \\ 
18aaxmvzj & 18:58:38.72 & +46:02:07.2 & 18.0 & 15.3-21.0 & --- & --- & --- & SOB,2 & 339 & --- & Y & --- & AT2016dta, G, Gx, MOT \\ 
18abloocr & 18:58:52.52  & +23:13:53.2  & 8.8 & 16.0-20.7 & $1.07\pm0.15$ & 935 & 10.8 & 3 & 350 & --- & --- & --- & AT2020sav, G \\ 
19aazxgcq & 19:00:15.06 & +42:32:41.8 & 16.5 & 16.1-20.5 & --- & --- & --- & 1 & 21 & --- & Y & --- & G \\ 
19abbwljz & 19:11:42.82 & +50:05:58.3 & 17.4 & 19.4-20.7 & --- & --- & --- & SOB & 38 & --- & --- & --- & AT2019ikt \\ 
18aakhgsc & 19:12:35.55 & +50:34:30.5 & 17.5 & 14.5-20.6 & $1.5\pm0.25$ & 667 & 11.4 & 2 & 363 & --- & --- & --- & G, KIC \\ 
19abpoypm & 19:14:51.00 & -11:32:35.2 & -10.4 & 15.7-20.0 & $1.06\pm0.16$ & 943 & 10.1 & 2 & 222 & --- & --- & --- & AT2020dta, G, MOT \\ 
18abiuwmf & 19:18:14.34 & +68:03:53.5 & 22.6 & 17.6-20.9 & $1.12\pm0.29$ & 894 & 11.1 & SOB,3 & 327 & --- & --- & --- & G \\ 
18abtivdb & 19:21:44.22 & +42:04:41.1 & 12.6 & 15.7-19.9 & --- & --- & --- & 4 & 340 & --- & --- & --- & G \\ 
19abagvei & 19:22:33.95 & +38:54:34.1 & 11.1 & 17.0-20.6 & --- & --- & --- & SOB,1 & 54 & Y & --- & --- & --- \\ 
20aaymduk & 19:23:15.52  & +25:17:23.1  & 4.8 & 17.5-19.7 & --- & --- & --- & SOB & 32 & --- & --- & --- & --- \\ 
18ablwwuk & 19:24:15.73 & +31:47:46.8 & 7.6 & 17.6-21.0 & --- & --- & --- & 4 & 106 & --- & --- & --- & G \\ 
18abrreod & 19:25:17.47  & +08:39:21.7  & -3.5 & 17.6-19.3 & --- & --- & --- & 3 & 365 & --- & --- & --- & AT2016iev, G \\ 
19abrkgih & 19:27:25.15 & -03:38:32.6 & -9.7 & 17.6-20.0 & --- & --- & --- & SOB & 43 & --- & --- & --- & --- \\ 
19aaxugwk & 19:28:22.35 & +55:32:01.1 & 17.2 & 13.7-21.0 & --- & --- & --- & SOB & 102 & --- & --- & SP & AT2019hau \\ 
19abfxdje & 19:29:08.09 & +08:18:38.7 & -4.5 & 14.9-20.3 & --- & --- & --- & SOB & 48 & --- & --- & --- & --- \\ 
18aazfdxy & 19:30:33.61 & +12:09:07.5 & -3.0 & 16.4-20.7 & $0.76\pm0.17$ & 1322 & 10.1 & 5 & 364 & --- & --- & --- & G \\ 
18aavghjc & 19:32:42.76  & +16:23:34.7  & -1.4 & 18.7-19.2 & --- & --- & --- & 1 & 26 & --- & --- & --- & --- \\ 
18abhnyca & 19:34:41.72 & +59:11:42.1 & 17.8 & 16.0-19.2 & $0.49\pm0.02$ & 2041 & 7.6 & 3 & 261 & --- & --- & --- & G \\ 
18acabezl & 19:35:29.07 & -01:10:59.0 & -10.3 & 18.6-19.4 & --- & --- & --- & SOB,1 & 287 & --- & --- & --- & AT2019pgk \\ 
18abudxbf & 19:36:08.62 & -06:37:41.0 & -12.9 & 17.2-19.7 & --- & --- & --- & 3 & 363 & --- & --- & --- & G \\ 
18abbkoer & 19:38:53.16 & +21:26:41.0 & -0.2 & 16.7-20.1 & --- & --- & --- & 3 & 338 & --- & --- & --- & G \\ 
18aboqywg & 19:39:22.49  & +66:53:47.7  & 20.3 & 15.6-19.6 & $1.08\pm0.09$ & 926 & 9.8 & 9 & 363 & --- & --- & --- & G, Gx, MOT  \\ 
18aaxdlbl & 19:41:25.00 & +15:22:54.3 & -3.7 & 15.2-18.9 & $4.64\pm0.06$ & 216 & 12.2 & 15 & 364 & --- & --- & --- & G \\ 
18aceglmy & 19:49:47.62 & +03:13:04.9 & -11.4 & 14.2-20.1 & $2.19\pm0.24$ & 457 & 11.8 & SOB,1 & 363 & --- & --- & --- & G \\ 
18abjfmwd & 19:49:55.19 & +45:53:50.8 & 9.9 & 15.8-20.2 & $1.13\pm0.13$ & 885 & 10.4 & 4 & 364 & --- & --- & --- & G, MOT \\ 
19abnnuse & 19:50:06.75 & +29:23:45.4 & 1.6 & 14.8-19.9 & $1.42\pm0.35$ & 705 & 10.7 & SOB,1 & 296 & --- & --- & --- & AT2019ndp, G \\ 
19aaxnqju & 19:50:19.73 & +03:26:29.8 & -11.4 & 17.7-19.6 & --- & --- & --- & SOB & 87 & --- & --- & --- & --- \\ 
18abcwcew & 19:50:21.10 & +41:52:35.2 & 7.8 & 16.7-20.9 & --- & --- & --- & 5 & 357 & --- & --- & --- & AT2018hni, 2M, G \\ 
18abucctv & 19:51:34.46  & +36:52:33.9  & 5.1 & 17.7-19.0 & --- & --- & --- & 1 & 25 & --- & --- & --- & KIC \\ 
19acfduug & 19:53:08.40  & -21:37:04.8  & -22.8 & 15.9-19.0 & --- & --- & --- & SOB & 50 & --- & --- & --- & AT2019yon \\ 
19abexjaf & 19:53:21.60 & +18:10:50.4 & -4.8 & 17.0-20.8 & --- & --- & --- & 1 & 36 & --- & --- & --- & AT2019kwx, G, MOT \\ 
19acawioe & 19:55:02.02 & +02:53:31.1 & -12.7 & 17.4-19.1 & --- & --- & --- & 2 & 273 & --- & --- & --- & G \\ 
18aaxluzd & 19:55:38.89  & +29:07:51.2  & 0.4 & 18.4-20.4 & --- & --- & --- & SOB,2 & 364 & --- & --- & --- & 2M \\ 
18abgjkgg & 20:00:02.31 & +58:08:41.4 & 14.4 & 18.3-20.7 & --- & --- & --- & 6 & 321 & --- & --- & --- & G \\ 
18abvazji & 20:02:03.25 & +54:47:30.1 & 12.5 & 15.7-20.9 & --- & --- & --- & SOB & 72 & --- & --- & --- & G \\ 
18abiwaxv & 20:04:23.11 & +44:20:29.8 & 6.9 & 15.5-20.8 & $0.89\pm0.24$ & 1130 & 10.5 & 3 & 328 & --- & --- & --- & G \\ 
19abrrkma & 20:05:06.68 & +16:20:18.6 & -8.1 & 16.4-20.1 & --- & --- & --- & 1 & 311 & --- & --- & --- & 2M, G \\ 
18abvwdss & 20:06:28.58 & +56:29:12.8 & 12.8 & 17.9-20.3 & --- & --- & --- & 4 & 343 & --- & --- & --- & G, MOT \\ 
18abmqumx & 20:08:27.91 & -07:09:10.8 & -20.4 & 16.0-19.4 & $0.62\pm0.12$ & 1626 & 8.4 & 3 & 359 & --- & Y & --- & G \\ 
19abpnahz & 20:12:25.09 & +49:43:21.2 & 8.6 & 17.1-20.6 & --- & --- & --- & 1 & 33 & --- & --- & --- & G \\ 
18acbwowf & 20:13:24.79 & +61:03:06.0 & 14.4 & 19.2-20.7 & --- & --- & --- & 2 & 343 & --- & --- & --- & AT2018hpk, PS15beo \\ 
18aavzlcg & 20:13:36.56  & +23:42:59.4  & -5.9 & 16.5-19.9 & --- & --- & --- & SOB,1 & 365 & --- & --- & --- & 2M, G \\
19abyvauj & 20:15:08.25 & +20:40:31.1 & -7.8 & 16.9-19.1 & --- & --- & --- & SOB & 36 & --- & --- & --- & G \\
19aawmbkz & 20:15:41.79 & +40:25:34.5 & 3.0 & 17.4-20.0 & $1.36\pm0.42$ & 737 & 10.6 & 2 & 353 & Y & --- & --- & G \\ 
19abgsssu & 20:16:16.71 & +18:22:25.4 & -9.3 & 14.8-20.3 & --- & --- & --- & SOB & 68 & --- & --- & --- & --- \\ 
18abnvugd & 20:16:28.18 & +16:53:31.2 & -10.2 & 18.1-20.9 & --- & --- & --- & 4 & 93 & --- & --- & --- & G \\ 
18abxyvdr & 20:20:16.12 & +24:51:31.0 & -6.5 & 17.7-19.0 & --- & --- & --- & SOB,1 & 87 & --- & --- & --- & AT2017chr \\ 
18acfbpwo & 20:22:37.67 & -19:05:33.8 & -28.4 & 18.5-19.9 & --- & --- & --- & 2 & 106 & --- & Y & --- & G \\ 
18abiknzw & 20:28:00.41 & +23:27:00.4 & -8.8 & 17.3-20.3 & --- & --- & --- & 7 & 363 & --- & --- & --- & G \\ 
18abktuub & 20:28:36.09 & +19:58:43.6 & -10.9 & 18.4-20.8 & --- & --- & --- & 6 & 358 & --- & --- & --- & G \\ 
19abjjtwi & 20:31:15.45 & +77:38:52.3 & 21.4 & 17.4-20.0 & --- & --- & --- & SOB & 361 & Y & --- & --- & AT2016fgw \\ 
18abgshhd & 20:34:59.64 & +25:55:29.8 & -8.6 & 17.7-20.3 & --- & --- & --- & SOB,3 & 361 & --- & --- & --- & G \\ 
18abwnokc & 20:35:29.77 & +06:36:52.6 & -19.6 & 16.4-20.7 & --- & --- & --- & 3 & 339 & --- & --- & --- & AT2016dyj, G, MOT \\ 
18absnrmp & 20:38:24.11 & +17:42:43.2 & -14.1 & 16.5-19.6 & --- & --- & --- & 7 & 357 & --- & --- & --- & AT2018juk, G, MOT \\ 
18acxgvqt & 20:39:37.57 & +21:39:06.0 & -12.0 & 15.3-19.5 & --- & --- & --- & SOB,1 & 347 & --- & --- & --- & 2M, G \\ 
18acswkeb & 20:40:07.13 & +26:00:31.7 & -9.5 & 17.0-19.4 & --- & --- & --- & SOB,1 & 277 & Y & --- & --- & AT2017evg, G \\ 
18aavqlll & 20:42:12.85 & +36:57:40.2 & -3.2 & 15.1-20.7 & $1.84\pm0.19$ & 544 & 12.0 & 2 & 364 & Y & --- & --- & G \\ 
18abhpycx & 20:42:56.96 & +23:52:46.4 & -11.3 & 17.8-20.3 & --- & --- & --- & SOB & 363 & --- & --- & --- & G \\ 
18abiklxj & 20:46:27.93 & +24:22:18.6 & -11.7 & 14.9-20.6 & $2.05\pm0.21$ & 488 & 12.2 & 6 & 363 & --- & --- & --- & G, MOT \\ 
18abtovmq & 20:46:28.68  & +22:34:15.0  & -12.7 & 16.7-20.7 & --- & --- & --- & 4 & 363 & --- & --- & --- & G \\ 
18ablvynr & 20:47:41.01 & +20:59:14.6 & -13.9 & 17.8-21.0 & --- & --- & --- & 3 & 361 & --- & --- & --- & G \\ 
18ablvxfm & 20:49:23.29 & +18:35:08.8 & -15.7 & 15.6-20.6 & $0.8\pm0.06$ & 1250 & 10.1 & 9 & 363 & --- & --- & --- & G \\ 
18abuktbx & 20:51:58.58 & +15:36:21.8 & -17.9 & 17.8-20.8 & --- & --- & --- & 7 & 363 & --- & --- & --- & G, MOT \\ 
18abnymoa & 20:53:55.98 & +34:14:09.4 & -6.8 & 17.3-19.3 & --- & --- & --- & SOB,2 & 363 & --- & --- & --- & G, MOT \\ 
18abvcdtl & 20:54:08.19 & -19:40:26.9 & -35.5 & 16.6-20.4 & --- & --- & --- & 5 & 178 & --- & Y & --- & AT2019dpb, G \\ 
19abfffna & 20:59:34.44 & +48:44:16.6 & 1.8 & 17.6-21.0 & --- & --- & --- & SOB & 82 & --- & --- & --- & AT2019lfr  \\ 
18abortbw & 21:01:40.43 & +21:57:31.8 & -15.9 & 16.9-19.8 & --- & --- & --- & 3 & 330 & --- & --- & --- & AT2016ern, MOT \\ 
18absnqxs & 21:03:34.58  & +23:41:28.3  & -15.1 & 15.0-20.4 & $0.94\pm0.15$ & 1064 & 10.3 & 3 & 356 & Y & --- & --- & 2M, G, Gx, MOT  \\ 
19aayvpky & 21:09:29.19 & -20:43:52.2 & -39.3 & 16.7-20.1 & --- & --- & --- & 4 & 162 & --- & Y & --- & AT2018hiv, G \\ 
18abastzb & 21:14:00.82 & +14:18:49.3 & -22.9 & 16.1-20.7 & $0.9\pm0.3$ & 1111 & 10.5 & 3 & 363 & --- & --- & --- & AT2017ecy, G \\ 
18abnbija & 21:15:43.18 & +29:30:14.8 & -13.4 & 18.5-20.5 & --- & --- & --- & 2 & 89 & --- & --- & --- & AT2019hif, G \\ 
18abktupp & 21:16:58.91  & +26:23:21.5  & -15.7 & 16.3-20.7 & $0.51\pm0.15$ & 1961 & 9.2 & 4 & 363 & --- & --- & --- & AT2019uot, G, Gx  \\ 
19abhpnxh & 21:19:19.43 & +41:12:35.2 & -5.9 & 17.3-20.7 & $1.77\pm0.45$ & 564 & 11.9 & HL & 299 & --- & --- & --- & AT2019uij, G \\ 
18abccncz & 21:20:29.94 & +21:17:33.2 & -19.7 & 18.8-20.4 & --- & --- & --- & 5 & 360 & Y & --- & --- & G \\ 
18acauhtv & 21:29:01.85 & +38:58:32.4 & -8.8 & 16.0-20.6 & --- & --- & --- & 2 & 341 & --- & --- & --- & G, MOT \\ 
19acfixfe & 21:31:50.81  & +49:14:01.7  & -1.7 & 17.4-19.2 & --- & --- & --- & 10 & 255 & Y & --- & K & AT2019weg, G \\ 
18acswtgb & 21:34:21.05 & -02:16:38.0 & -36.8 & 18.5-20.3 & $1.68\pm0.31$ & 595 & 11.4 & 10 & 360 & Y & --- & K & G, Gx \\ 
18abmmzzz & 21:35:54.80 & +23:36:44.2 & -20.7 & 16.4-20.9 & --- & --- & --- & 4 & 197 & Y & Y & --- & G \\ 
18abwnzaf & 21:38:06.62 & +32:19:42.1 & -14.8 & 17.4-20.1 & $0.66\pm0.18$ & 1515 & 9.2 & SOB,1 & 351 & --- & --- & --- & G \\ 
18abcosdv & 21:38:09.25 & +21:28:23.8 & -22.6 & 16.3-21.0 & --- & --- & --- & 3 & 215 & Y & Y & --- & AT2020nlv, G, MOT \\ 
17aaaajfc & 21:42:44.19 & +47:09:13.5 & -4.4 & 16.2-19.1 & --- & --- & --- & SOB & 41 & Y & --- & --- & --- \\ 
18abmegwl & 21:43:51.22 & +31:53:24.1 & -16.0 & 16.9-19.4 & --- & --- & --- & 4 & 194 & --- & --- & --- & G, MOT \\ 
18abadjtv & 21:44:26.45  & +22:20:24.6  & -23.0 & 15.2-20.2 & $0.96\pm0.09$ & 1042 & 10.1 & 1 & 334 & Y & Y & --- & 2M, G, Gx  \\ 
18abmnrmj & 21:53:44.29 & +32:14:46.4 & -17.1 & 15.9-20.5 & --- & --- & --- & SOB,1 & 361 & --- & Y & --- & G \\ 
18abkhxsl & 21:56:27.82 & +37:02:04.9 & -13.8 & 17.5-20.6 & $0.94\pm0.3$ & 1064 & 10.5 & 3 & 358 & --- & --- & --- & G \\ 
18abudxrk & 21:56:30.45 & -03:19:57.7 & -42.0 & 17.7-20.9 & --- & --- & --- & 3 & 193 & Y & Y & --- & G \\ 
18abcoxqa & 21:56:36.35 & +19:32:41.5 & -27.0 & 16.4-20.7 & $2.09\pm0.27$ & 478 & 12.3 & 4 & 342 & Y & Y & --- & G \\ 
19acewtjx & 21:57:34.12 & +36:25:01.4 & -14.4 & 16.0-19.3 & --- & --- & --- & SOB & 54 & --- & --- & --- & AT2019ufm, ASASSN-19zl \\ 
18abmarqq & 21:59:06.52 & +35:24:35.0 & -15.4 & 16.6-21.0 & --- & --- & --- & 3 & 345 & --- & Y & --- & AT2020jcg, G \\ 
18abjuxmo & 22:03:28.22 & +30:56:36.4 & -19.4 & 15.6-20.6 & $1.01\pm0.12$ & 990 & 10.7 & 6 & 345 & Y & Y & --- & G \\ 
19abfalqj & 22:08:46.35 & +47:50:18.0 & -6.6 & 18.3-19.5 & --- & --- & --- & SOB & 122 & --- & --- & --- & --- \\ 
19aaahelh & 22:10:51.35  & +41:21:30.5  & -12.1 & 17.6-19.1 & --- & --- & --- & 3 & 346 & Y & --- & --- & --- \\ 
18abuemed & 22:11:54.93  & +43:09:16.8  & -10.7 & 17.4-20.8 & $0.77\pm0.2$ & 1299 & 10.2 & SOB,2 & 364 & Y & --- & --- & G \\ 
18abnygtg & 22:13:44.02 & +17:32:51.6 & -31.2 & 15.8-21.0 & $1.74\pm0.46$ & 574 & 12.2 & 1 & 55 & Y & Y & --- & G \\ 
18abvxaxf & 22:16:21.96  & +70:54:15.9  & 11.8 & 15.9-19.9 & --- & --- & --- & SOB,2 & 178 & Y & --- & --- & 2M, G, MOT  \\ 
18abscevg & 22:18:16.26  & +40:53:05.6  & -13.3 & 18.8-20.6 & --- & --- & --- & 4 & 363 & Y & --- & --- & G, Gx  \\ 
17aaburxr & 22:27:47.28  & +68:43:35.6  & 9.4 & 17.4-20.2 & --- & --- & --- & 4 & 278 & Y & --- & --- & G \\ 
19abuqlzt & 22:31:37.43 & +40:29:43.3 & -15.0 & 16.7-20.6 & $1.58\pm0.42$ & 633 & 11.6 & SOB & 92 & --- & --- & --- & G \\ 
18abaqipz & 22:33:36.03 & +51:37:39.2 & -5.6 & 16.9-20.2 & $0.5\pm0.07$ & 2015 & 8.7 & 13 & 362 & --- & --- & --- & G \\ 
18abcnnfj & 22:35:21.95 & +33:02:43.8 & -21.7 & 16.4-21.0 & $3.01\pm0.19$ & 332 & 13.4 & 5 & 360 & Y & Y & --- & AT2016dyv, 2M, G \\ 
19acarefd & 22:35:59.48 & +46:27:51.1 & -10.3 & 17.1-20.1 & --- & --- & --- & SOB & 60 & --- & --- & --- & AT2019rdl \\ 
18abdeymj & 22:42:53.41 & +17:25:37.9 & -35.6 & 17.4-19.1 & --- & --- & --- & 2 & 283 & Y & Y & --- & AT2019xl, G \\ 
18abgtfuk & 22:44:05.83 & +43:45:32.2 & -13.3 & 16.2-19.3 & $0.72\pm0.05$ & 1389 & 8.6 & 13 & 363 & --- & --- & --- & 2M, G, MOT \\ 
18abzacuc & 22:46:13.75 & +57:32:36.1 & -1.3 & 17.8-19.7 & --- & --- & --- & SOB & 40 & --- & --- & --- & --- \\ 
17aaajxht & 22:51:21.68  & +66:33:46.8  & 6.4 & 15.3-18.4 & $1.36\pm0.02$ & 735 & 9.1 & 16 & 362 & --- & --- & --- & 2M, G \\ 
17aabpkjj & 22:55:03.35 & +54:38:19.3 & -4.5 & 17.9-20.1 & --- & --- & --- & 5 & 198 & --- & --- & --- & G \\ 
17aabutck & 23:01:48.71  & +41:19:05.0  & -17.0 & 17.2-19.2 & --- & --- & --- & SOB & 125 & --- & --- & --- & G, MOT  \\ 
18abumcmz & 23:15:37.05 & +50:06:16.6 & -9.9 & 17.7-20.9 & --- & --- & --- & 7 & 331 & --- & --- & --- & AT2019tjl, G \\ 
18abfyvlc & 23:15:52.35 & +27:10:37.2 & -31.0 & 14.9-20.8 & --- & --- & --- & SOB,1 & 150 & Y & Y & --- & G, Gx \\ 
18abnzbxh & 23:19:36.07 & +36:46:59.6 & -22.5 & 13.8-20.5 & $2.46\pm0.1$ & 407 & 12.4 & 3 & 362 & --- & Y & --- & G \\ 
19aceiizg & 23:20:18.51 & +49:19:46.0 & -10.9 & 15.2-20.0 & --- & --- & --- & SOB,1 & 49 & --- & --- & --- & AT2019slj, ASASSN-19za \\ 
17aabulyc & 23:26:19.41 & +28:26:49.7 & -30.8 & 16.8-20.3 & --- & --- & --- & 3 & 364 & Y & Y & --- & G, Gx \\ 
19abajklf & 23:31:26.79 & +47:36:52.5 & -13.1 & 16.4-18.6 & --- & --- & --- & 1 & 126 & Y & --- & --- & AT2018eoh \\ 
19abzpkss & 23:40:37.56 & +42:35:27.7 & -18.4 & 14.3-18.3 & --- & --- & --- & SOB & 69 & Y & --- & --- & AT2019qqk \\ 
18abncfvc & 23:43:24.80 & +42:08:40.7 & -19.0 & 14.8-19.7 & --- & --- & --- & 1 & 158 & Y & Y & --- & G \\ 
19abritjp & 23:43:31.32 & +47:43:16.2 & -13.6 & 17.6-20.6 & --- & --- & --- & 1 & 19 & --- & --- & --- & AT2018kmh, G \\ 
19aavnjma & 23:47:59.95 & +76:39:02.4 & 14.2 & 14.2-20.7 & $2.34\pm0.27$ & 427 & 12.6 & SOB & 134 & --- & --- & --- & AT2019rsl, G \\ 
\enddata 
\tablenotetext{a}{K=Keck, SP=SPRAT, F=Floyds} 
\tablenotetext{b}{MOT=MASTEROT, G=Gaia, Gx=GALEX, 2M=2MASS} 
\end{deluxetable} 
\end{longrotatetable}

\clearpage
\begin{figure}
\plotone{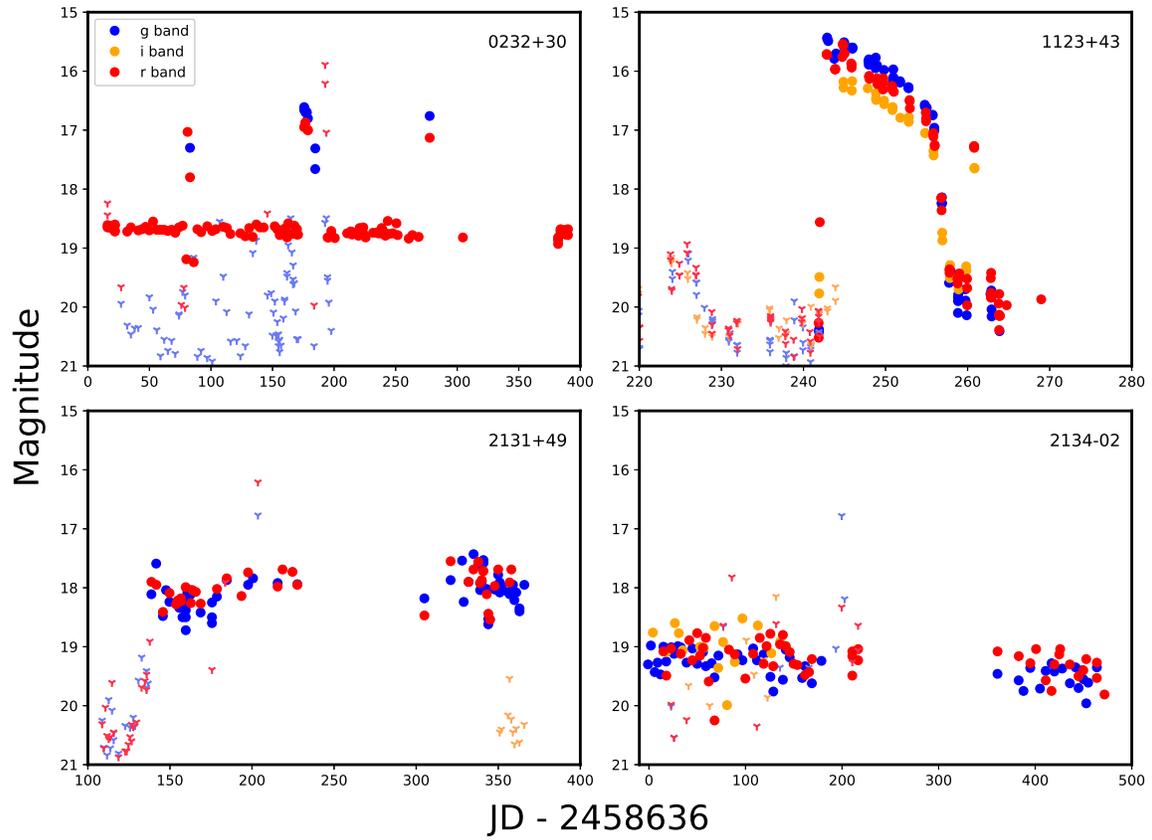}
\caption{Examples of ZTF light curves of CV candidates from Table 2. Filled blue, red and yellow circles are magnitudes from $g, r,i$ filters, while light symbols are upper limits on those nights. Upper left is an example of a typical short outburst cycle dwarf nova, upper right is a typical dwarf nova superoutburst light curve, and bottom light curves correspond to the spectra shown in Figure 2.}
\end{figure}

\clearpage
\begin{figure}
  \includegraphics[width=7in]{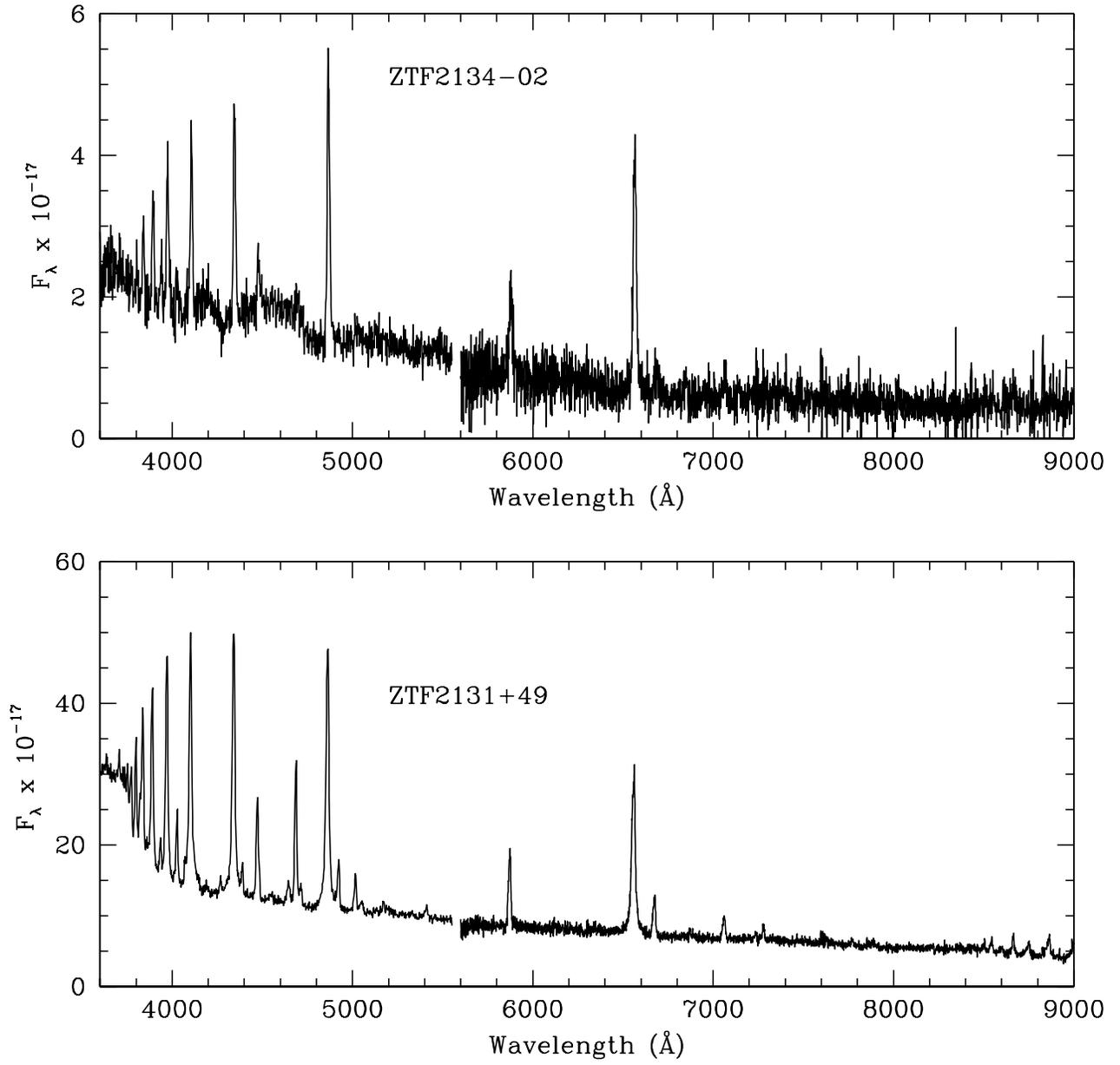}
  \caption{Blue and red spectra from Keck
  confirming these 2 objects as CVs. The vertical axis is F$_{\lambda}$ in units
  of 10$^{-17}$ ergs cm$^{-2}$ s$^{-1}$ \AA$^{-1}$. Note the strong \ion{He}{2}4686 line
  in ZTF2131+49 that could indicate a magnetic white dwarf.}
  \end{figure}
  
\clearpage\begin{figure}
    \centering
    \includegraphics[width=7in]{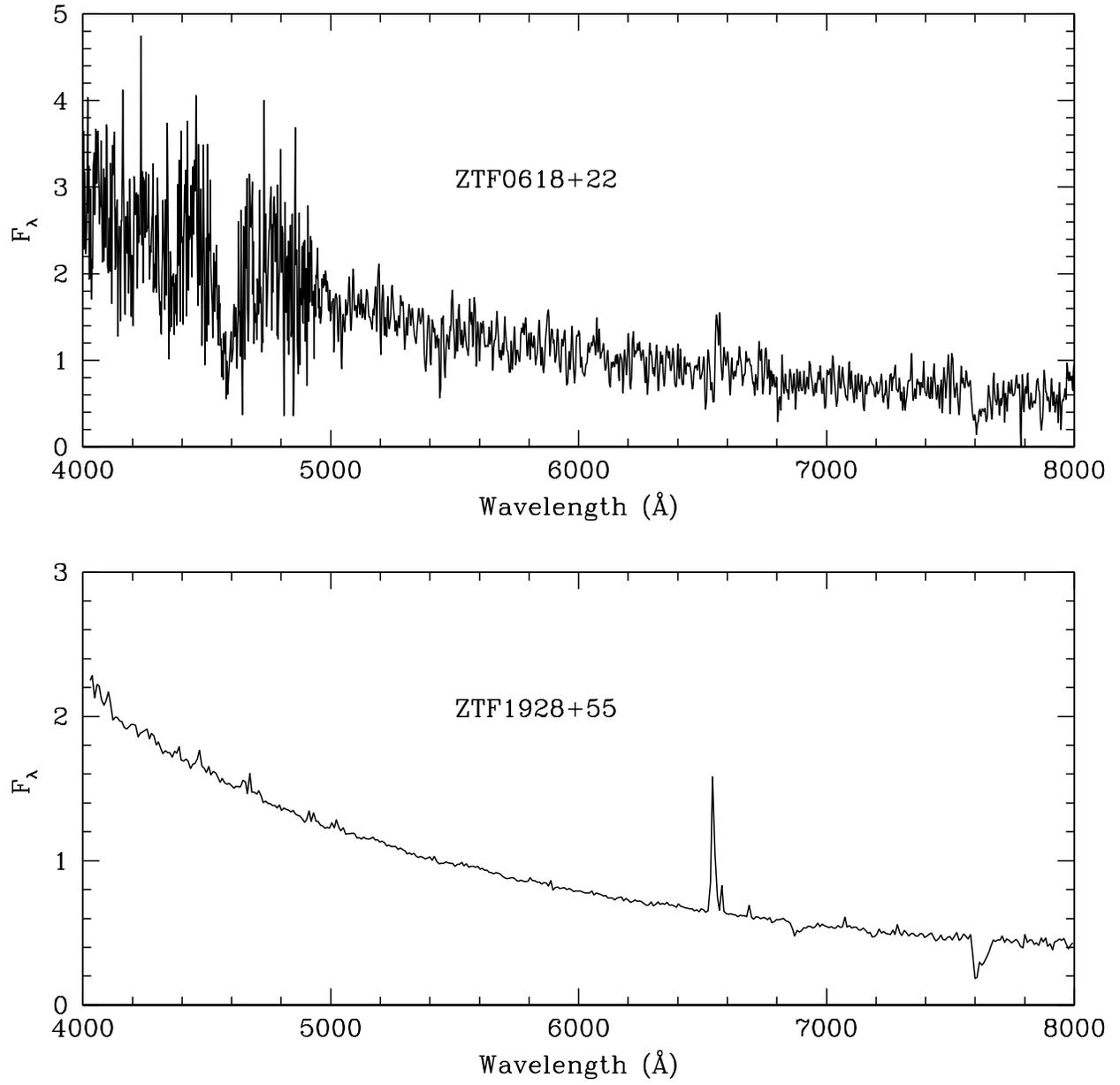}
    \caption{Low resolution spectra. Top: Floyds spectrum of ZTF0618+22 showing H$\alpha$ emission. Bottom: SPRAT spectrum of ZTF1928+55 with H, He and \ion{He}{2} emission. The vertical axis is F$_{\lambda}$ in units
  of 10$^{-17}$ ergs cm$^{-2}$ s$^{-1}$ \AA$^{-1}$.}
    \label{fig:my_label}
\end{figure}

\clearpage
\begin{figure}
    \centering
    \includegraphics[width=7in]{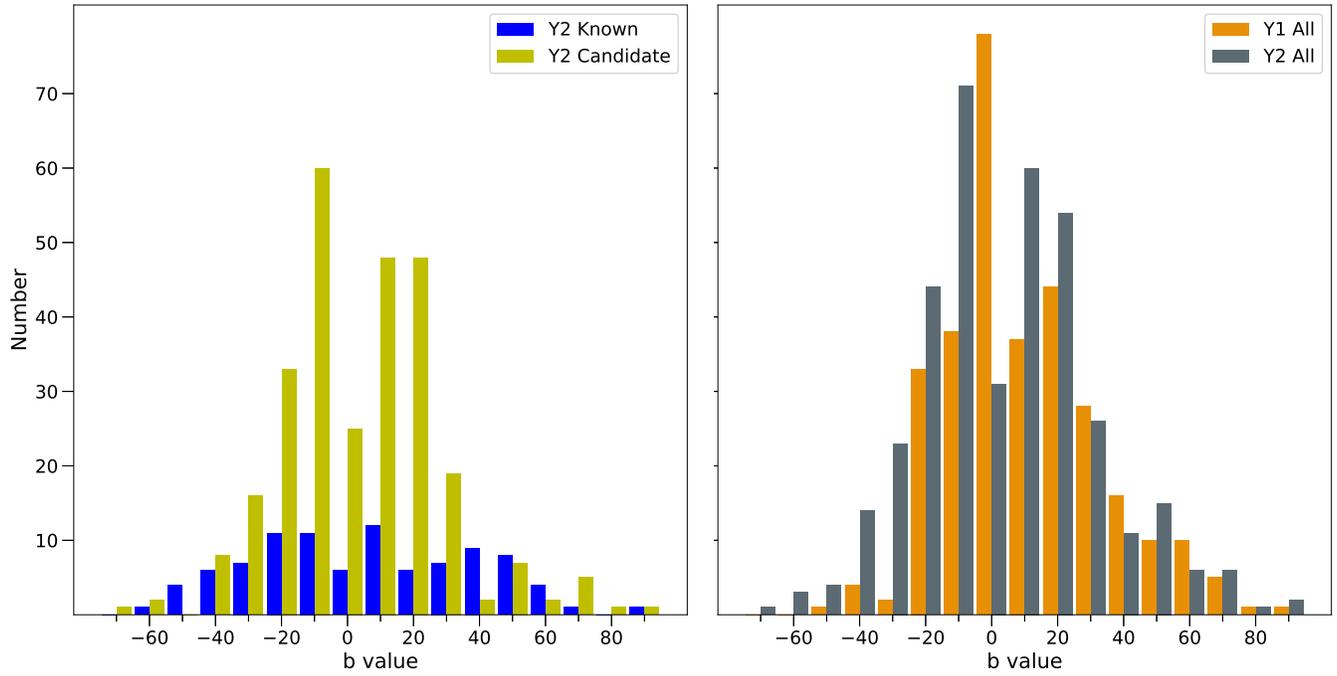}
    \caption{Left:The number of systems in Tables 1-2 as a function of galactic latitude (in 10 deg bins) during the 2nd year. Right: All CVs from Tables 1 and 2 in the first year compared to the second year. }
\end{figure}

\clearpage
\begin{figure}
    \centering
    \includegraphics[width=7in]{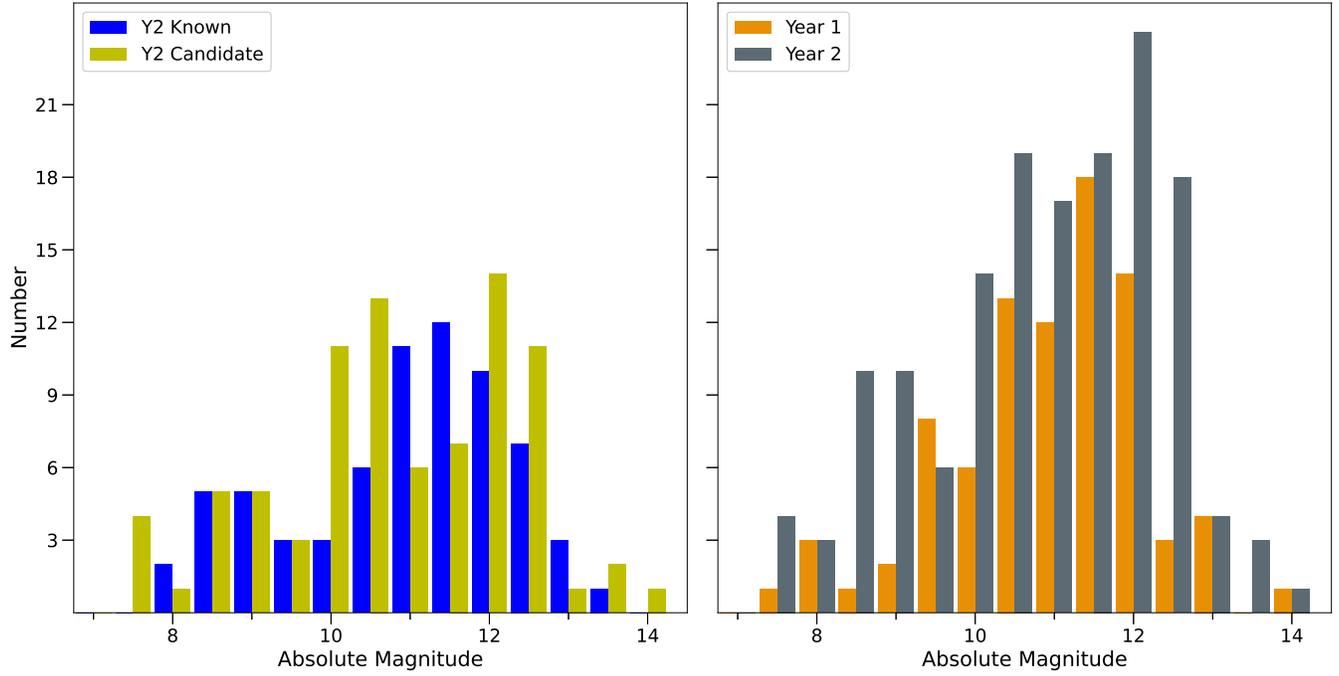}
    \caption{Left: The number of systems in Tables 1 and 2 for the 2nd year as a function of their absolute magnitude (in 0.5 mag bins) with distances from available Gaia parallaxes. Due to the 5$\sigma$ upper limits on the magnitudes for the fainter sources, the numbers are only bright limits on the absolute magnitude at quiescence. Right: Comparison of total known and candidates for year 1 and year 2.}
\end{figure}

\clearpage
\begin{figure}
    \centering
    \includegraphics[width=7in]{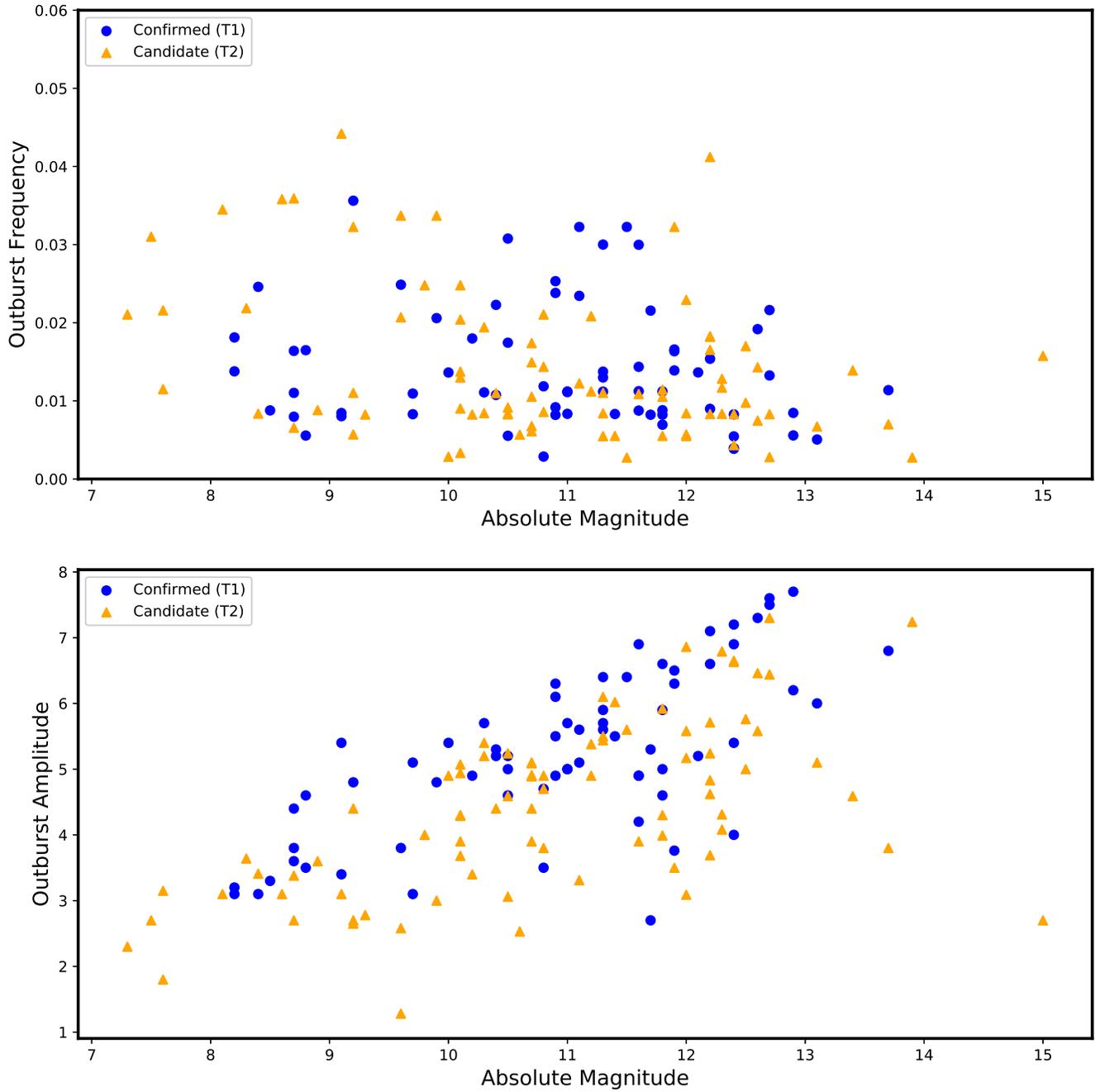}
    \caption{Plots of Absolute magnitude versus Outburst Frequency (top)
and Absolute Magnitude versus Outburst Amplitude (bottom) for the known dwarf novae in Table 1 (solid dots) and those showing dwarf nova type outbursts in Table 2 
(triangles).}
\end{figure}

\clearpage
\begin{figure}
    \centering
    \includegraphics[width=7in]{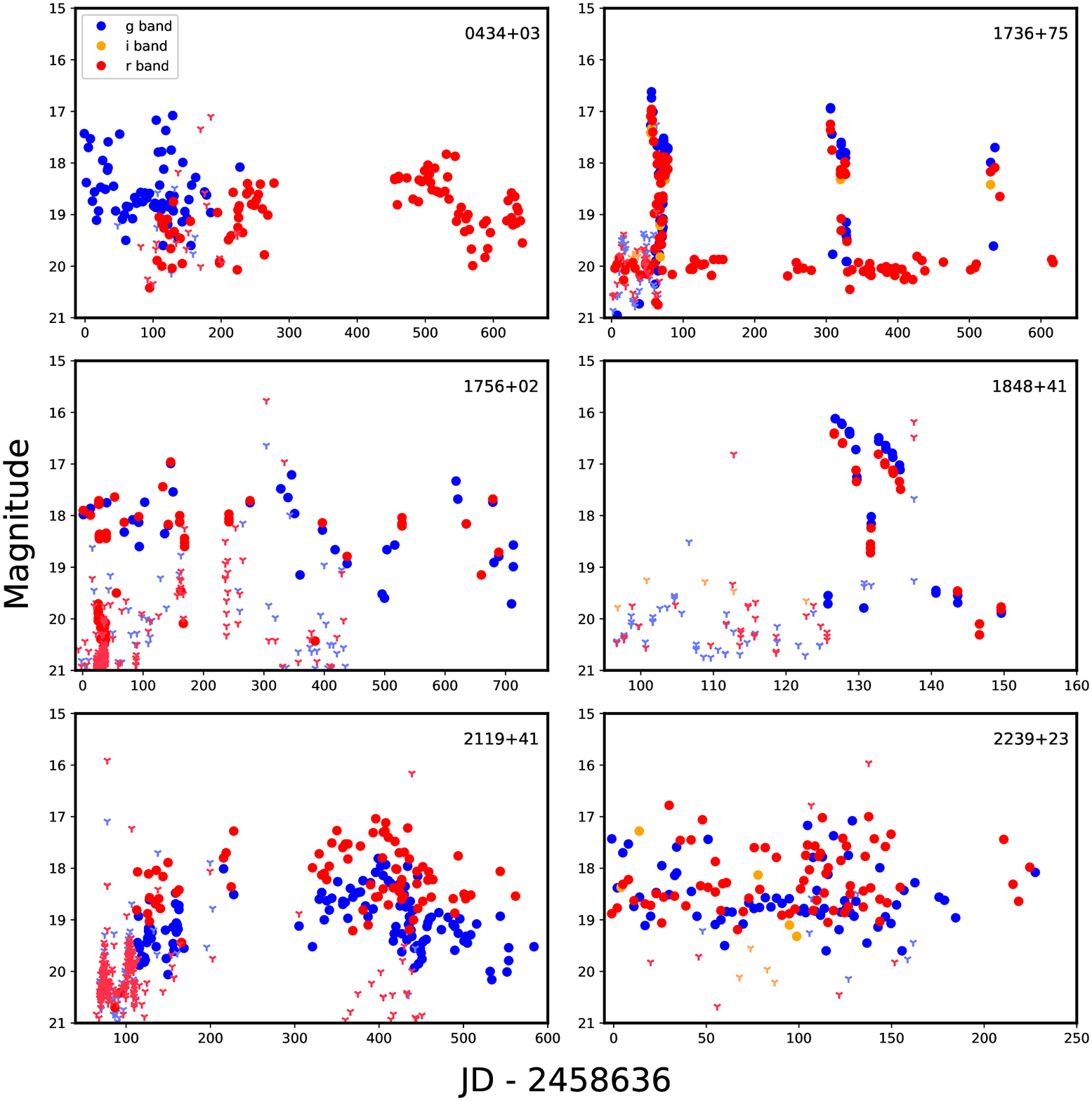}
    \caption{ZTF light curves of some peculiar objects discussed in text. Symbols are the same
    as in Figure 1.}
\end{figure}

\end{document}